%% file: main.tex
\documentclass[aps,onecolumn,manuscript,amsmath,amssymb,prb]{revtex}
\usepackage{graphicx}
\usepackage{graphics}
\usepackage{color}
\usepackage{rotating}

\def \i{\`{\i}~}

\begin{document}

\title{
Some issues concerning Large-Eddy Simulation of inertial particle dispersion in
turbulent bounded flows
}

\author{
C. Marchioli$^1$, M.V. Salvetti$^2$ and A. Soldati$^{1,*}$
\footnote[1]{$^{*}$Author to whom correspondence should be addressed.
Also affiliated with Department of Fluid
Mechanics, CISM, 33100 Udine, Italy. Electronic Mail: 
soldati@uniud.it.}
}
\address{
$^1$ Dept. Energy Technology and
Centro Interdipartimentale di Fluidodinamica e Idraulica,
University of Udine, 33100 Udine -- Italy\\
$^2$ Dept. Aerospace Engineering, University of Pisa, 56100 Pisa -- Italy\\
}

\maketitle

\newcommand{\mbf}{\mathbf}
\newcommand{\Pec}{\rm P_c}
\newcommand{\Pe}{\rm Pe}

\flushbottom
\topmargin  -16mm
\textwidth  140mm
\textheight 240mm
\columnsep   10mm
\parindent    4mm


\section*{\small\bf ABSTRACT}

The problem of an accurate  Eulerian-Lagrangian
modeling of inertial particle dispersion in Large
Eddy Simulation (LES) of turbulent wall-bounded flows
is addressed. We run Direct Numerical Simulation (DNS)
for turbulent channel flow at $Re_{\tau}=150$ and $300$
and corresponding {\em a priori} and {\em a posteriori} LES
on differently coarse grids. We then tracked swarms of
different inertia particles and we examined the influence of filtering
and of Sub-Grid Scale (SGS) modeling for the fluid
phase on particle velocity and
concentration statistics.
We also focused on how particle preferential segregation
is predicted by LES.
Results show that even ``well-resolved'' LES is unable to reproduce
the physics as demonstrated by DNS, both for particle accumulation
at the wall and for particle preferential segregation.
Inaccurate prediction is observed for the entire range of particles
considered in this study, even when the particle response time is much larger
than the flow timescales not resolved in LES.
Both {\em a priori} and {\em a posteriori}
tests indicate that recovering the level of
fluid and particle velocity fluctuations
is not enough to have accurate prediction
of near-wall accumulation and local segregation.
This may suggest that reintroducing the correct amount
of higher-order moments of the velocity fluctuations is
also a key point for SGS closure models for the particle
equation.
Another important issue is the presence of possible flow
Reynolds number effects on particle dispersion.
Our results show that, in small Reynolds number turbulence and in the case
of heavy particles, the shear fluid velocity is a suitable scaling
parameter to quantify these effects.
%
%

\section{Introduction}
\label{intro}

The dispersion of inertial particles in
turbulent flows is characterized by macroscopic phenomena such as
non-homogeneous distribution, large-scale clustering and preferential
concentration due to the inertial bias between the denser
particles and the lighter surrounding fluid. \cite{hit2,ms02}
In homogeneous isotropic turbulence, \cite{hit2,hit1,fs06} clustering
and preferential concentration may be crucial in determining collision
frequency, breakage efficiency, agglomeration and
reaction rates.
Focusing on turbulent boundary layer, we observe that,
besides controlling particle interaction rates, these
phenomena strongly influence settling, deposition and entrainment.\cite{s05}

Both Direct Numerical Simulation (DNS) \cite{ms02,mgss03} and
Large-Eddy Simulation (LES) \cite{uo96,ws96,k06}
together with Lagrangian Particle Tracking (LPT) have been widely used to
investigate and quantify these macroscopic phenomena, for instance in
channel \cite{ms02,ws96,k06} or in pipe flows. \cite{mgss03,uo96}
DNS-based Eulerian-Lagrangian studies are widely used for investigating the physics
of particle-turbulence interactions, whereas LES has yet
to demonstrate its full capabilities in predicting
correctly particle-turbulence statistics \cite{kv05}
and macroscopic segregation phenomena. \cite{fs06,k06}
In previous studies on particle dispersion in turbulent
bounded flows, \cite{kv05,k06} LES methods were found inadequate
to predict local segregation phenomena which eventually control
macroscopic fluxes.
This inadequacy was attributed to the filtering of the flow Sub-Grid Scales
(SGS).
To elaborate, in LES-based Eulerian-Lagrangian simulations of particle dispersion
a {\em sub-grid} error is introduced in the particle equation
since only
the filtered fluid velocity is available;
this approximation adds to the {\em modeling} error
which is intrinsic to the SGS modeling. \cite{kv05}
Similar to what is done for the flow field,
a way to model the effects of the filtered
SGS velocity fluctuations on particle motion should be identified.\cite{k06}

Among previous LES applications to gas-solid turbulent
flows, \cite{uo96,armenio} the SGS
velocity fluctuations were neglected
based on the assumption that the particle
response time was large compared to the smallest timescale
resolved in the LES.\cite{armenio}
It was later demonstrated that this assumption
leads to a certain degree of inaccuracy on the prediction of
particle velocity statistics and concentration.
In particular, the results obtained by Kuerten and Vreman \cite{kv05}
and by Kuerten \cite{k06} for turbulent dispersion of heavy particles
in channel flow have shown that, due to both sub-grid and modeling errors,
LES underestimates the tendency of particles
to move towards the wall by the effect of the turbulence
(turbophoretic effect\cite{r83}).
To circumvent this problem, a closure model for the particle equation
of motion based on filter inversion
by approximate deconvolution was proposed to recover the
influence of the filtered scales in turbulent channel flow  \cite{k06}
and in homogeneous turbulent shear flow. \cite{sm05}

An effort was also provided to establish criteria according to which
the SGS modeling for particles could be judged necessary or not. In particular,
F\'{e}vrier {\em et~al.} \cite{Fevrier2005} 
have shown that LES filtering has an effect on particle motion which
depends on the ratio of the particle size to the
filtered spatial scales.
Fede and Simonin \cite{fs06} have studied the influence
of sub-grid fluid velocity fluctuations on particle dispersion, preferential
concentration, and inter-particle collisions in homogeneous
isotropic turbulence.
For single particle statistics
such as turbulent dispersion, Fede and Simonin \cite{fs06} confirm
that an explicit accounting of sub-grid fluid turbulence on
particle transport is not required when the particle response
time is much larger than the cut-off timescale of the sub-grid
velocities. However, they show also that
accumulation and collision phenomena are strongly influenced
by sub-grid fluid turbulence even when the particle
response time is up to ${\cal{O}}(10)$ times the Kolmogorov time
scale.

Aim of the present study is precisely to build on
the work of Kuerten and Vreman \cite{kv05}
extending the analysis of Fede and Simonin \cite{fs06} on the sub-grid
turbulence effects on particle accumulation (neglecting inter-particle collisions)
to turbulent channel flow, which presents the additional complexity
of a solid wall and of turbulence strong anisotropy and non-homogeneity.
The analysis is grounded on a systematic
investigation on the importance of SGS effects on particle motion
for different particle inertia and under different flow conditions.
The specific objectives of the present study can be summarized as follows:
{\em (i)} Create a database for particle-laden
turbulent channel flow in which different values of the flow
Reynolds number and of the particle response time are considered,
showing in this paper particle velocity and concentration statistics
reported from DNS/LES computed at
shear Reynolds number up to 300 and grid resolution
up to $256^3$ grid points.
{\em (ii)} Use this database to investigate on the importance of the
SGS velocity fluctuations in predicting the statistical properties
of the dispersion process. We will focus on
the effects due to changes in particle inertia (obtained by
tuning of the particle size with respect to the filtered spatial scales)
or in grid resolution as well as on possible flow Reynolds
number scaling properties of particle preferential concentration.
{\em (iii)} Provide subsidies for the development of new
SGS closure models for the equations of particle motion.

The paper is organized as follows. Problem statement, governing equations and numerical
methodology required for the simulations are presented in Sec. \ref{meth}. 
Sec. \ref{results} is devoted to the analysis and discussion of relevant statistics obtained from
simulations where particle trajectiores are computed from DNS, filtered
DNS in {\em a priori} tests and LES in {\em a posteriori} tests. In this case
the discussion will be focused on the quantification of sub-grid and modeling
errors on particle velocity and concentration statistics as well as on particle
preferential distribution, thus extending the analysis of Kuerten and Vreman.\cite{kv05}
Finally, conclusions and future developments are drawn in Sec. \ref{conclusions}.

\section{Physical Problem and Numerical Methodology}
\label{meth}

\subsection{Particle-laden turbulent channel flow}

The flow into which particles are introduced is a turbulent
channel flow of gas.
In the present study, we consider air (assumed
to be incompressible and Newtonian)
with density $\rho = 1.3~kg~m^{-3}$ and kinematic
viscosity $\nu = 15.7{\times}10^{-6}~m^{2}~s^{-1}$.
The governing balance equations for the fluid (in
dimensionless form) read as:
\begin{equation}
\label{cont}
\frac{\partial{u_i}}{\partial{x_i}} = 0~,
\end{equation}
\begin{equation}
\label{ns}
\frac{\partial{u_i}}{\partial{t}} = -u_{j}\frac{\partial{u_i}}{\partial{x_j}} +
\frac{1}{Re}\frac{\partial^{2}{u_i}}{\partial{x_j}^2} - \frac{\partial{p}}{\partial{x_i}}
+ \delta_{1,i}~,
\end{equation}
where $u_i$ is the $i^{th}$ component of the dimensionless velocity
vector, $p$ is the fluctuating kinematic pressure, $\delta_{1,i}$
is the mean dimensionless pressure gradient that drives the flow
and $Re_{\tau} = u_{\tau}h/\nu$ is the shear Reynolds number
based on the shear (or friction) velocity, $u_{\tau}$,
and on the half channel height, $h$.
The shear velocity is defined as $u_{\tau} = (\tau_{w}/\rho)^{1/2}$,
where $\tau_{w}$ is the mean shear stress at the wall.
All variables considered in this study are reported in dimensionless
form, represented by the superscript + (which has been dropped from Eqns.
(\ref{cont}) and (\ref{ns}) for ease of reading) and expressed in wall units. Wall
units are obtained combining $u_{\tau}$, $\nu$ and $\rho$.
%
%

In LES, the standard Continuity and Navier-Stokes equations are smoothed
with a filter function of width $\Delta$. Accordingly, all flow
variables are decomposed into a resolved (large-scale) part and a residual
(sub-grid scale) part as
${\bf u}({\bf x},t)={\bf \bar{u}}({\bf x},t)+\delta{\bf u}({\bf x},t)$.
The filtered Continuity and Navier-Stokes equations for the resolved scales are then:

\begin{eqnarray}
\label{cont-les}
\frac{\partial \bar{u}_i}{\partial x_j} = 0~,
\end{eqnarray}
\vspace{-0.6cm}
\begin{equation}
\label{ns-les}
\frac{\partial{\bar{u}_i}}{\partial{t}} = -\bar{u}_{j}\frac{\partial{\bar{u}_i}}{\partial{x_j}} +
\frac{1}{Re}\frac{\partial^{2}{\bar{u}_i}}{\partial{x_j}^2} - \frac{\partial{\bar{p}}}{\partial{x_i}}
+ \delta_{1,i} - \frac{\partial{\tau_{ij}}}{\partial{x_j}}~,
\end{equation}
where ${\tau_{ij}}={\overline{u_i u_j}} - \bar{u}_i\bar{u}_j$ represents the sub-grid scale
stress tensor. The large-eddy dynamics is closed once a model for ${\tau_{ij}}$ is provided.
In the present study, the dynamic SGS model of Germano et al. \cite{germano} has been applied.

The reference geometry consists of two infinite
flat parallel walls: the origin of the coordinate system is located
at the center of the channel and the $x-$, $y-$ and $z-$ axes point
in the streamwise, spanwise and wall-normal directions respectively
(see Fig.~\ref{f-chandomain}).
Periodic boundary conditions are imposed on the fluid velocity field
in $x$ and $y$, no-slip boundary conditions
are imposed at the walls.
The calculations were performed
on a computational domain of size
$4 \pi h \times 2 \pi h \times 2 h$
in $x$, $y$ and $z$ respectively.
%
%

Particles with density $\rho_p=1000~kg~m^{-3}$ are injected
into the flow at concentration low enough to
consider dilute system conditions.
The motion of
particles is described by a set of ordinary differential
equations for particle velocity and position.
For particles much heavier than the fluid ($\rho_{p}/\rho \gg 1$)
Elghobashi and Truesdell \cite{et92} have shown that the most
significant forces are Stokes drag and buoyancy
and that Basset force can be neglected being
an order of magnitude smaller.
In the present simulations, the aim is to minimize the
number of degrees of freedom by keeping the simulation setting
as simplified as possible; thus
the effect of gravity has also been
neglected. With the above assumptions, a
simplified version of the Basset-Boussinesq-Oseen
equation \cite{cst98} is obtained.
In vector form:
\begin{eqnarray}
\label{partpos}
\frac{d{\bf x}}{dt} = {\bf v}~,
\end{eqnarray}
\vspace{-0.6cm}
\begin{eqnarray}
\label{partvel}
\frac{d{\bf v}}{dt} =
-\frac{3}{4}\frac{C_D}{d_p}\left( \frac{\rho}{\rho_p} \right) |{\bf v} - {\bf u}|({\bf v} - {\bf u})~,
\end{eqnarray}
where ${\bf x}$ is particle position,
${\bf v}$ is particle velocity, and
${\bf u}$ is fluid velocity
at particle position.
The Stokes drag coefficient is computed as
$C_D = \frac{24}{Re_p}(1+0.15Re_{p}^{0.687})$
where
$Re_p = d_{p}|{\bf v} - {\bf u}|/\nu$
is the particle Reynolds number.
The correction for $C_D$ is necessary when $Re_p$ does not remain
small.

\subsection{DNS and LES methodology}

In this study both DNS and LES have been applied to the fully-developed
channel flow.
In both cases, the governing equations are discretized
using a pseudo-spectral method
based on transforming the field variables into wavenumber space, using Fourier
representations for the periodic
streamwise and spanwise directions and a Chebyshev representation for the
wall-normal (non-homogeneous) direction.
A two level, explicit
Adams-Bashforth scheme for the non-linear terms,
and an implicit
Crank-Nicolson method for the viscous terms are employed
for time advancement.
Further details of the method have been published previously.
\cite{PB_POF_96}

DNS calculations have been performed using the parallel (MPI) version of the
flow solver.
Two values of the shear Reynolds number have been considered
in this study: $Re_{\tau} = 150$ ($Re^l_{\tau}$ hereinafter)
based on the shear velocity $u_{\tau}^l = 0.11775~m~s^{-1}$,
and $Re_{\tau} = 300$ ($Re^h_{\tau}$ hereinafter) based
on the shear velocity $u_{\tau}^h = 0.2355~m~s^{-1}$.
The corresponding average (bulk) Reynolds numbers are
$Re^l_{b} = u_b^l h / \nu = 1900$, where $u_b^l \simeq 1.49~m~s^{-1}$
is the average (bulk) velocity; and $Re^h_{b} = u_b^h h / \nu = 4200$,
where $u_b^h \simeq 3.3~m~s^{-1}$, respectively.
The size of the computational domain in wall units
is $1885\times942\times300$ for the $Re^l_{\tau}$ simulations and
$3770\times1885\times600$ for the $Re^h_{\tau}$ simulations.
The computational domain has been discretized in physical space
with $128\times128\times129$ grid points (corresponding to $128\times128$
Fourier modes and to 129 Chebyshev coefficients in the wavenumber space)
for the $Re^l_{\tau}$ simulations and with $256\times256\times257$
grid points (corresponding to $256\times256$ Fourier modes and to
257 Chebyshev coefficients in the wavenumber space) for the
$Re^h_{\tau}$ simulations in order to keep the grid spacing
fixed. This is the minimum number of grid points required in each direction
to ensure that the grid spacing is always smaller than the smallest
flow scale
and that the limitations imposed
by the point-particle approach are satisfied. \cite{footnote}

LES calculations have been performed using the serial version of the
pseudo-spectral flow solver on the same computational domain.
Two computational
grids have been considered: a {\em coarse} grid made of $32\times32\times65$
nodes and a {\em fine} grid made of $64\times64\times65$ nodes. Only the
lower value, $Re^l_{\tau}$, of the shear Reynolds number has been considered.

The complete set of DNS/LES simulations is summarized in Table \ref{fluid-sim}.

\subsection{Filtering for {\em a priori} tests}
\label{filtering}

In the {\em a priori} tests the Lagrangian tracking of particles is
carried out starting from the filtered velocity field,
$\bar{{\bf u}}$,
obtained through explicit filtering of the DNS velocity by means
of either a cut-off or a top-hat filter. Both filters are applied
in the homogeneous streamwise and spanwise directions in the
wave number space:
\begin{eqnarray}
\label{cut-off}
\bar{u}_i ({\bf x},t) = FT^{-1} \left\{
\begin{array}{ll}
G(\kappa_1) \cdot G(\kappa_2) \cdot
\hat{u}_i (\kappa_1,\kappa_2,z,t)~~~~~~{\text{if}}~~| \kappa_j | \le |\kappa_c|
~~{\text{with}}~~j=1,2~,\\
0 \hspace{5.75cm} \text{otherwise~.}
\end{array}
\right.
\end{eqnarray}
where $FT$ is the 2D Fourier Transform, $\kappa_c= \pi / \Delta$ is the
cutoff wave number ($\Delta$ being the filter width in the physical
space), $\hat{u}_i (\kappa_1,\kappa_2,z,t)$
is the Fourier transform of the fluid velocity field, namely
$\hat{u}_i (\kappa_1,\kappa_2,z,t)=FT [\bar{u}_i ({\bf x},t) ]$ and $G(\kappa_i)$ is
the filter transfer function:
\begin{eqnarray}
\label{transfer-function}
G(\kappa_j) = \left\{
\begin{array}{ll}
1~~~~~~~~~~~~~~~~~~~{\text{for the cut-off filter~,}}\\
\frac{sin(\kappa_j \Delta/2)}{\kappa_j \Delta/2} ~~~~~~~~~\text{for the top-hat filter~.}
\end{array}
\right.
\end{eqnarray}
Three different filter widths have been considered at $Re_{\tau}^l$,
corresponding to a grid Coarsening Factor (CF in Table \ref{fluid-sim})
in each homogeneous direction of 2, 4 and 8 with respect to DNS.
In the wall-normal direction data are not filtered,
since often in LES the wall-normal resolution is DNS-like. \cite{pope-faq}


\subsection{Lagrangian particle tracking}

To calculate particle trajectories in the flow
field, we have coupled a Lagrangian tracking routine
with the DNS/LES flow solver.
The routine solves for Eqns. (\ref{partvel}) and (\ref{partpos})
using $6^{th}$-order
Lagrangian polynomials to interpolate fluid velocities at
particle position; with this velocity the equations
of particle motion are advanced in time
using a $4^{th}$-order Runge-Kutta scheme.
The timestep size used for particle tracking was
chosen to be equal to the timestep size used for
the fluid, $\delta t^+ = 0.045$; the total tracking time
was, for each particle set, $t^+ = 1200$ in the
{\em a priori} tests and $t^+ = 1800$ in the
{\em a posteriori} tests.
These simulation times are
not long enough to achieve a statistically
steady state for the particle concentration. \cite{test-case}

Particles, which are assumed pointwise, rigid and spherical,
are injected into the flow at concentration low enough to neglect
particle collisions.
The effect of particles onto the turbulent field
is also neglected (one-way coupling assumption).
At the beginning of the simulation, particles
are distributed homogeneously over
the computational domain and their initial
velocity is set equal to that of the fluid
at the particle initial position.
Periodic boundary conditions are imposed on particles
moving outside the computational domain in the
homogeneous directions,
perfectly-elastic collisions at the smooth walls were assumed
when the particle center was at a distance lower than one
particle radius from the wall.

For the simulations presented here,
large samples of $10^{5}$ particles, characterized by
different response times, were considered
for each value of $Re_{\tau}$. The response time
is defined as $\tau_{p} = \rho_{p}d^{2}_{p}/18\mu$
where $\mu$ is the fluid dynamic viscosity: when
the particle response time is made
dimensionless using wall variables,
the Stokes number for each particle
set is obtained as $St=\tau_p^+=\tau_p/\tau_f$
where $\tau_f=\nu/u_{\tau}^2$ is the
viscous timescale of the flow.
Tables~\ref{part} and ~\ref{part300} show all the parameters
of the particles injected into the flow field.
We remark here that, for the present channel flow configuration
at $Re_{\tau}^l$, the non-dimensional value
of the Kolmogorov timescale, $\tau_K^+$, ranges
from 2 wall units at the wall to 13
wall units at the channel centerline. \cite{jot}
Hence, if
we rescale the particle response times given
in Table~\ref{part} using the local value of
$\tau_K^+$ near the centerline, where the flow conditions
are closer to homogeneous and isotropic, we obtain
Stokes numbers that vary from $10^{-2}$ to $10$ and
fall in the lower range of values considered by Fede and
Simonin. \cite{fs06}

As a further remark, we wish to add that the characteristic timescale of the
flow changes depending on the specific value of the shear
Reynolds number, namely on the specific value of the shear
velocity. In the present case, we have
$\tau_f^l = \nu / \left( {u_{\tau}^{l}} \right)^{2} = 1.13 \cdot 10^{-3}~s$
for the $Re_{\tau}^l$ simulations and
$\tau_f^h = \nu / \left( {u_{\tau}^{h}} \right)^2 = 2.83 \cdot 10^{-4}~s$
for the $Re_{\tau}^h$ simulations. Elaborating, we find
that the same value of the Stokes number corresponds
to different (dimensional) values of the particle response time
according to the following expression:

\begin{equation}
\label{scalingtaup}
St^l=St^h \rightarrow
\frac{\tau_p^l}{\tau_f^l}=\frac{\tau_p^h}{\tau_f^h}
\rightarrow
\frac{\tau_p^l}{\tau_p^h}=\frac{\tau_f^l}{\tau_f^h}=
\left( \frac{u_{\tau}^h}{u_{\tau}^l} \right)^2 =
\left( \frac{Re_{\tau}^h}{Re_{\tau}^l} \right)^2=4~,
\end{equation}
where $St^h$ and $St^l$ represent the particle Stokes number
in the $Re_{\tau}^l$ simulation and in the $Re_{\tau}^l$
simulation, respectively.

%
%
%
%
%

\section{Results}
\label{results}

\subsection{Particle distribution in {\em a priori} LES at $Re_\tau$=150}
\label{a-priori}

In this Section, we will discuss the influence of filtering on particle distribution
by showing the velocity statistics and the concentration profiles for particles
dispersed in {\em a priori} LES flow fields, i.e. filtered DNS fields. We will also discuss
filtering effects on local particle preferential segregation using macroscopic
segregation parameters.
As described in Sec. 2.3, the cut-off and the top-hat filters have
been used; the first one provides a sharp separation between resolved
and non-resolved scales and can be considered the filter corresponding
to a coarse spectral simulation, in which no explicit filtering
is applied. Conversely, the top-hat filter is a smooth filter and,
thus, it subtracts a significant amount of energy from the resolved
scales. \cite{Sagaut} For each filter, three different
filter widths have been considered.
Fig. \ref{spectrum-les}
sketches the effect of these filter widths on the
one-dimensional (streamwise) frequency spectrum, $E(\omega)$, \cite{pope}
computed for the $Re_{\tau}=150$ flow.
Since particle dynamics in the viscous sublayer is controlled
by flow structures with timescale $\tau_f$ around 25 and considering
that this timescale corresponds to the circulation
time of the turbulence structures in the buffer layer
($5 < z^+ < 30$), \cite{pms05} we show the energy spectrum
at the $z^+=25$ location. The cut-off frequencies corresponding
to each filter width are indicated as
$\omega_{{\text{cut-off}}}^{{\text{CF=2}}}$,
$\omega_{{\text{cut-off}}}^{{\text{CF=4}}}$
and
$\omega_{{\text{cut-off}}}^{{\text{CF=8}}}$ in increasing order.
Also shown (dot-dashed lines) are the estimated
response frequencies which characterize
each particle set considered in the {\em a priori} tests,
these frequencies being proportional to $1/\tau_p$.
Areas filled with patterns below the energy
profile represent the relative amount of energy
removed by each filter width: larger filter widths
prevent particles from being exposed to ever-increasing
turbulent frequencies, namely to smaller and smaller flow scales
which can modify significantly their local behavior,
dispersion and segregation. Inaccurate estimation of these
processes due to filtering will bring sub-grid errors into
subsequent particle motion.

Fig. \ref{nfi} shows the particle root mean square (rms)
velocity fluctuations obtained in the
{\em a priori} tests with cut-off filter for the $St=1$,
the $St=5$ and the $St=25$ particles, respectively.
The reference values
obtained injecting the particles in the DNS flow velocity fields are
also reported. Specifically,
the streamwise and wall-normal rms components
are shown in Figs. \ref{nfi}(a-c) and Figs. \ref{nfi}(d-f).
All profiles were
obtained averaging in time (from $t^+=450$ to $t^+=1200$) and space (over the homogeneous
directions).
It is apparent that filtering the fluid velocity has
a large impact on the turbulent velocity
fluctuations. As expected, particle velocity fluctuations are reduced
in particular for the larger filter widths corresponding to
coarser LES grids. This is consequence of the well known decrease of the flow
velocity fluctuations due to filtering as felt by the particles,
even if in a different measure depending on their
inertia. Note, however, that the effect of filtering is significant
also on particles having characteristic
response frequencies much lower than those removed by the filters
(e.g. the $St=25$ particles). Finally, for the cut-off filter,
underestimation of the particle fluctuations is a pure effect
of the elimination of the SGS scales, since no energy is subtracted
from the resolved ones. The results obtained with the top-hat filter
(not shown here for brevity) are qualitatively similar, although,
for a given filter width,
underestimation of the particle fluctuations is, as expected, larger for the top-hat
filter than for the cut-off one.
The reduction of the wall-normal velocity
fluctuations near the wall for the {\em a priori} LES, shown
in Figs. \ref{nfi}(d-f), is worth noting because it corresponds
to a reduction of particle turbophoretic drift (namely, particle
migration to the wall in turbulent boundary layers) and, in turn,
to a reduction of particle accumulation in the near-wall region.
\cite{kv05} This is also shown in
Fig. \ref{a-priori-filtering} in which the near wall instantaneous
particle concentration
obtained in {\em a priori} LES is compared to the DNS one for different
filter widths and particle inertia.
Concentration profiles shown here are taken at time $t^+=1200$: as mentioned,
the particle tracking in {\em a
priori} LES was not carried out long enough to reach a
statistically-steady particle
concentration at the wall. However, we checked many different time
instants and, although the concentration values change, the trend is
always the same as that shown in Fig. \ref{a-priori-filtering}. It
appears that, consistently with the results of Kuerten and Vreman, \cite{kv05}
filtering leads to a significant underestimation of the wall particle
concentration, for all filter types and widths and for
all particle sets considered in this study.

Finally, in Fig. \ref{poisson-dns-a-priori-les} the particle
segregation parameter, $\Sigma_p$, is plotted versus the particle Stokes
number in two different regions of the channel: the channel centerline,
where $\Sigma_p$ has been computed in a fluid slab 10 wall unit thick
centered at $z^+=150$, and in the near-wall region, where $\Sigma_p$ has
been computed in the viscous sublayer ($0 \le z^+ \le 5$). The segregation
parameter (or maximum deviation from randomness)\cite{re01} is
calculated as $(\sigma - \sigma_{Poisson})/m$,
where $\sigma$ and $\sigma_{Poisson}$ represent respectively
the standard deviations for the particle number density distribution
and the Poisson distribution.
The particle number density distribution is computed on a grid containing
$N_{cell}$ cells of volume $\Omega_{cell}$ covering the entire computational
domain.
The parameter $m$ is the mean number of particles in one
cell for a random uniform particle distribution.
\cite{re01,poisson}
The drawback of this method is the dependence of $\Sigma_p$ on the cell
size. To avoid this problem, we computed the particle number density distribution
for several values of $\Omega_{cell}$ and we kept only the largest value
of $\Sigma_p$. \cite{Fevrier2005}
First, as found in previous studies, \cite{pms05}
for this shear Reynolds number a peak of $\Sigma_p$ occurs for
$St \simeq 25$ and preferential concentration falls off on either side of this
{\em optimum} value. As shown for instance in Fig.
\ref{a-priori-filtering}, $St=25$ particles are thus the most responsive to
the near-wall turbulent structures. When an explicit filter is applied,
particle segregation is
underpredicted severely in all considered cases, especially near the
wall. Note that this underestimation is significant also for the
smallest filter width, for which the reduction of particle
fluctuations was relatively small (see Fig. \ref{nfi}).

\subsection{Particle distribution in {\em a posteriori} LES at $Re_\tau$=150}
\label{a-posteriori}

In this Section, we will discuss the behavior of particles
dispersed in LES flow fields. Two different LES grids have been
used, as shown in Table I. In
these {\em a posteriori} tests, different sources of errors are
present in addition to the filtering effects discussed in Sec. 3.1,
viz. the errors due to
{\em (i)} the SGS modeling for the fluid phase,
{\em (ii)} the numerical discretization of the fluid governing equations and
{\em (iii)} the interpolation in the Lagrangian particle tracking.
For the used pseudo-spectral
discretization the numerical error should plausibly be negligible. As
for interpolation, a $6^{th}$-order interpolation scheme is used. Although we
did not carry out a sensitivity study, the analysis in Kuerten and Vreman
\cite{kv05} indicates that the
interpolation error should remain small, even if it may introduce an
additional smoothing. Thus, we believe that the main source of difference
with the {\em a priori} tests is represented by the SGS model closing
the governing equations for the fluid phase. As in the {\em a priori}
tests, no closure model is used in the equations of particle motion.

In order to asses the quality of the LES for the fluid part, Fig. \ref{rms-dns-les}
compares the streamwise and wall-normal rms of the fluid velocity
components obtained in LES to the reference DNS values. For the more
resolved LES, a good agreement with DNS is obtained and, hence, this
can be considered as a {\em well-resolved} LES for the fluid phase.
Conversely, in the
coarser LES significant errors are found in the prediction of the
fluid-phase velocity fluctuations and, thus, errors in the Lagrangian
particle tracking are anticipated. The effect of
the SGS modeling error is clearly visible if the values obtained for
the coarser grid are compared with those of the {\em a-priori} tests
in Fig. \ref{nfi} for a corresponding coarsening factor
(CF=4). Indeed, in the {\em a posteriori} LES, the introduction of the SGS
model tends to counteract the decrease of the fluid velocity
fluctuations due to filtering; in the coarser case this leads to an
overestimation of the rms of the streamwise and wall-normal velocity
components. This overestimation is a rather well-known behavior of
coarse LES, especially for the rms of the streamwise
component. Nonetheless, it is worth remarking that in actual LES the
fluid velocity fields in which the particles are dispersed are not
always characterized by a lack of fluctuations, as it happens
in the idealized context of {\em a priori} tests. As previously mentioned, the
dynamic eddy-viscosity model \cite{germano} was used to close the LES equations for
the fluid phase. We also carried out LES simulations with the
Smagorinsky model, but, as expected, the
results were generally less accurate than those obtained with the
dynamic SGS model for a fixed resolution: hence they are not shown or discussed here
for sake of brevity.

In Fig. \ref{rmspart-dns-les} the streamwise and wall-normal rms of
the different particle sets obtained in LES are compared with the
reference DNS data. A good agreement with DNS is obtained in
the more resolved LES for all the considered particle inertia, while
for the coarser simulation significant discrepancies are found. Note
that in the coarse case, the rms of the wall-normal velocity component
are overestimated for all the considered particle sets, as previously
observed also for the fluid phase.
In spite of the differences in the fluid and particle
velocity fluctuations observed in {\em a priori} and {\em a posteriori}
tests, the underestimation of particle concentration at the wall, already
observed in the {\em a priori} tests (see Sec. 3.1), is also
found in {\em a posteriori} LES,  for all considered
resolutions and particle sets. This is shown, for instance, by the
instantaneous particle
concentration profiles of
Fig. \ref{a-posteriori-filtering}. The same is for the
underprediction of the particle preferential concentration (see
Fig. \ref{poisson-dns-a-posteriori-les}). It is worth nothing that the
errors on the quantitative prediction of both particle segregation
and near-wall accumulation are large also for the well-resolved LES, in
which the level of fluid and particle
velocity fluctuations is rather well predicted. This indicates that, in order to obtain
acceptable predictions for near-wall accumulation and
particle segregation, the
reintroduction of the correct level of velocity fluctuations is not the only issue
to devise a closure model for the particle equations.
Finally, in the {\em a posteriori} LES the segregation parameter,
$\Sigma_p$, was also computed for $St=125$ particles (see
Fig. \ref{poisson-dns-a-posteriori-les}). For this set of particles,
the values obtained in both LES simulations are higher than those
computed in DNS.
In their {\em a priori} tests
for homogeneous and isotropic turbulence, Fede and Simonin \cite{fs06} found that for particles
having lower inertia than a given threshold value the effect of filtering was
to decrease the segregation parameter, while for particles of larger inertia the
segregation was conversely increased. From our results, this scenario seems to hold also
in {\em a posteriori} LES and in near wall turbulence.


\subsection{Influence of the Reynolds number on particle distribution}
\label{scaling}

{\em A priori} and {\em a posteriori} simulations have emphasized that the
importance of SGS velocity fluctuations in predicting the properties of particle
dispersion depends both on particle inertia, parametrized quantitatively by
the particle Stokes number, and on the spatial resolution of the Eulerian grid.
%
The results shown in previous sections, however, are relative to a turbulent
channel flow at $Re_{\tau}=150$, a rather small value compared to those
typical of LES applications. If higher values of $Re_{\tau}$ are to be
considered, then Reynolds number effects on particle dispersion may
become significant because the characteristic
length and time scales of the particle change
with respect to those of the fluid when the flow dynamics
change: in particular, the higher the Reynolds number the smaller the particle
response time for a given value of the Stokes number (see discussion in Sec. 2.4).
This point can be further elucidated considering Fig. \ref{spectrum}, where
the frequency spectrum, $E(\omega)$ computed
for the $Re_{\tau}^l$-DNS (already shown in Fig. \ref{spectrum-les})
is compared with the frequency spectrum computed
for the $Re_{\tau}^h$-DNS. The spectrum is computed at a wall-normal
distance $z^+=25$ consistently with the reasons discussed in Fig. \ref{spectrum-les}.
As done in that figure,
the characteristic response frequencies of the particles
are also shown.
It is apparent that, in the
$Re_{\tau}^h$-flow {\em (i)} the turbulent kinetic energy budget
is associated to a wider range of frequencies,
namely to smaller flow timescales with which the particles may
interact, and {\em (ii)} a given value of
frequency corresponds to higher values of
the turbulent kinetic energy.
In this case, applying cut-off frequencies like those shown in
Fig. \ref{spectrum-les} will remove
a wider spectrum of flow scales and a larger amount of turbulent
kinetic energy from the flow with respect to
the lower Reynolds number flow, thus magnifying the error-behavior
of the model in predicting particle segregation and wall accumulation.

In principle, these observations should lead to the conclusion that
SGS models must incorporate a dependency on the flow Reynolds number.
In fact, the need to include Reynolds number effects
should be assessed carefully being based on the knowledge
of how particle preferential concentration scales with $Re_{\tau}$.
Numerical investigations on the Reynolds number scaling properties
of the preferential concentration of heavy particles have been performed
in a synthetic turbulent advecting field by Olla \cite{o02}
and in homogeneous isotropic turbulence by
Collins and Keswani.\cite{ck04}
Here, we investigate on the same effect in turbulent channel flow.

To introduce our scaling argument, let us
rewrite Eq. (\ref{scalingtaup}) for the case of constant particle
response time (namely ${\tau_p^h}={\tau_p^l}$). We have:

\begin{equation}
\label{scalingSt}
\frac{St^h}{St^l}=
\frac{\tau_f^l}{\tau_f^h}=
\left( \frac{u_{\tau}^h}{u_{\tau}^l} \right)^2 =
\left( \frac{Re_{\tau}^h}{Re_{\tau}^l} \right)^2=4~.
\end{equation}

From Eq. (\ref{scalingSt})
we can conclude the following: if the shear velocity
is the proper scaling parameter to quantify the Reynolds number
effect on particle preferential concentration
then the statistical description of the $St^h=4$ particles behavior
in the $Re_{\tau}^h$-flow is expected to resemble that of the $St^l=1$
particles behavior in the $Re_{\tau}^l$-flow. Similarly,
scaling effects are expected to
couple the $St^h=20$ particles
to the $St^l=5$ particles
and the $St^h=100$ particles
to the $St^l=25$ particles,
respectively.
We thus expect that, for instance, the particle velocity fluctuations
should be proportional to the fluid velocity fluctuations within the
range of Reynolds number considered in this study, provided that
the Reynolds number effect on preferential concentration is scaled
properly.

Figure \ref{scaling-DNS} compares {\em vis-\`{a}-vis} the ratios between
the particle and the fluid velocity rms components, $v'_{i,rms}/u'_{i,rms}$,
normalized to wall variables using the shear velocity as the scaling parameter.
The non-dimensional distance from the wall is indicated as $z^+/H^+$
where $0 \le z^+ \le H^+$, $H^+$ being equal to either $Re_{\tau}^l$ or $Re_{\tau}^h$.
In each panel of Fig. \ref{scaling-DNS}, thick lines
are used for the $Re_{\tau}^l$-DNS whereas open symbols
refer to the $Re_{\tau}^h$-DNS. For ease of reading, $v'_{i,rms}/u'_{i,rms}$
profiles for the streamwise and wall-normal rms components are shifted by a factor
of 0.4, up and down respectively.
These statistics have been computed
averaging over a time window corresponding to the last $720$ wall
time units of the simulations
under statistically-developing conditions for the particle concentration.
Rms ratios have been computed for all particle sets,
yet comparison is made
only between the particle Stokes numbers matching the two
different Reynolds numbers according to Eq. (\ref{scalingSt}):
namely $St^l=1$ and
$St^h=4$ (Fig. \ref{scaling-DNS}a), $St^l=5$ and $St^h=20$ (Fig. \ref{scaling-DNS}b),
$St^l=25$ and $St^h=100$ (Fig. \ref{scaling-DNS}c).
%
%
%
It is apparent that the profiles, though a bit ragged, overlap quite well
even in the near-wall region, where discrepancies (possibly due to the
extension of the averaging time window) are limited to very
thin slabs inside the viscous sublayer,
thus supporting the validity of the
adopted scaling.

In Fig. \ref{poisson-dns-dns} particle segregation in the center
of the channel
(Fig. \ref{poisson-dns-dns}a) and in the near-wall region
(Fig. \ref{poisson-dns-dns}b) is quantified by the segregation
parameter $\Sigma_p$ for the two DNS simulations.
Black symbols represent the values of $\Sigma_p$ for the
five sets of particles considered in the $Re_{\tau}^l$-DNS, whereas
open symbols are used for the six sets of particles considered in
the $Re_{\tau}^h$-DNS.
Two observations can be made: first, lower segregation occurs
at higher Reynolds number for a given value of the particle Stokes number;
second, the degree of segregation is nearly same for
particle Stokes numbers and shear Reynolds numbers matching
the condition given in Eq. (\ref{scalingSt}), as
indicated by the dot-dashed lines with arrows.
This is particularly true in the near-wall region.

These results seem to indicate that particle
preferential concentration scales proportionally to the flow Reynolds
number and that the particle Stokes number, defined as particle timescale
normalized to wall variables (the shear fluid velocity being the scaling
parameter) may be used to characterize the coupling between particles and
fluid in the regime where particles preferentially concentrate.
These effects appear to be consistent with other observations,
most of which refer to the classical Kolmogorov scaling argument
\cite{ck04,yps06}
that predict statistical saturation at higher
Reynolds numbers than those considered here.

\clearpage

\section{Conclusions and Future Developments}
\label{conclusions}

In this paper, the problem of assessing an accurate Eulerian-Lagrangian
modeling of heavy particle dispersion in Large Eddy Simulation is
addressed. This problem is investigated in a systematic way by performing
DNS, {\em a priori} and {\em a posteriori} LES coupled
with Lagrangian particle tracking of fully developed channel flow, in
which different values of the flow Reynolds number and of the particle
response time have been considered. The accuracy in the prediction of the
particle velocity statistics, near wall accumulation and
preferential segregation are assessed through {\em vis-\`{a}-vis} comparison
against DNS data.

Consistently with the results of Kuerten and Vreman, \cite{kv05}
the effect of pure filtering in {\em a priori} tests is to decrease
the fluid velocity fluctuations and, in turn, the particle velocity
fluctuations, although by different amounts according to particle
inertia. This leads to a
severe underestimation of particle accumulation at the wall.
Extending the analysis to particle segregation, quantified by
a macroscopic indicator, we found that filtering leads to a significant
underestimation of particle preferential concentration.
In conclusion, it appears that a closure model is needed for the
particle equations. \cite{ws96,k06,kv05,sm05}
In {\em a posteriori} LES simulations, we have found that the SGS
dynamic model, exploited to close the problem for the fluid phase,
is able to reintroduce a correct level of fluid velocity fluctuations
when a rather fine grid (two times the DNS grid spacing in each direction)
is used; the
particle velocity fluctuations are also in good agreement
with those obtained in DNS. Conversely, significant discrepancies are
observed with respect to the DNS reference values when a coarser
resolution (typical of LES applications) is used.
We observe that the velocity fluctuations
of both phases are overestimated, in contrast with the {\em a priori}
tests.
Despite these differences, particle wall accumulation
and local segregation are always severely underestimated.
This indicates that the reintroduction of the
correct level of fluid and particle velocity fluctuations is not the only
issue for accurate SGS closure models for the particle equations,
which apparently is not enough to have an accurate prediction of near-wall
accumulation and local particle segregation. It may be argued that,
since these phenomena are governed by complex interactions between the
particles and the flow structures, the reitroduction of
the correct amount of higher order moments of the velocity fluctuations
for both phases is probably the key point to develop these models.
This could be achieved, for instance, using non-Gaussian stochastic
Lagrangian models based on Langevin-type equations. \cite{hanratty}

Another important feature of SGS models for particles
is that they are required to account for possible flow Reynolds
number effects on particle accumulation and segregation.
Albeit the Reynolds number range limitations of this study,
we have shown that scaling of statistics
seems to persist for the Reynolds numbers considered ($Re_{\tau}=150$
and $300$) and therefore it is possible to parametrize Reynolds number
effects simply by imposing a quadratic dependence of the particle
Stokes number, defined as the ratio of the particle response
time to the viscous timescale of the flow, on the
shear Reynolds number.
Finally, we are aware that the results shown here only cover
the lower range of Reynolds numbers typical of LES
applications: Hence, one future development (currently under way)
of this work
will be to investigate on the Reynolds number scaling
properties of particle segregation through DNS/LES of
turbulent channel flow at $Re_{\tau}=600$.

\section*{Acknowledgments}

The authors wish to thank A. Mannucci, L. Rigaux and L. Osmar
for their help in performing some of the simulations.
Support from PRIN (under Grant 2006098584$\_$004) and from HPC
Europa Transnational Access Program (under Grants 466 and 708)
are gratefully acknowledged.

\include{references}


\include{tables}

\include{figures}

\end{document}

%% file: references.tex
\bibliographystyle{plainnat}

%% file: tables.tex
%
%
\begin{table}[t]
\begin{minipage}[t]{15.0cm}
\begin{small}
\begin{center}
\begin{tabular}{c c c c}
$Re_{\tau}$ & DNS & {\em a-priori} LES & {\em a-posteriori} LES\\
\hline
 & & $64 \times 64 \times 129$ (CF=2) & $64 \times 64 \times 65$\\
150 (=$Re_{\tau}^l$) & $128 \times 128 \times 129$ & $32 \times 32 \times 129$ (CF=4) & $32 \times 32 \times 65$\\
 & & $16 \times 16 \times 129$ (CF=8) & --- \\
\hline
300 (=$Re_{\tau}^h$)& $256 \times 256 \times 257$ & --- & --- \\
\end{tabular}
\vspace{0.3cm}
\caption{\label{fluid-sim} Summary of the simulations.}
\end{center}
\end{small}
\end{minipage}
\end{table}
\vspace{-0.8cm}
%
%
\begin{table}[t]
\begin{minipage}[t]{15.0cm}
\begin{small}
\begin{center}
\begin{tabular}{c c c c c c}
$St^l=St |_{Re_{\tau}^l}$ & $\tau_{p}^l~(s)$ & $d_{p}^{+}$ & $d_{p}$ (${\mu}m$) & $V_{s}^{+}=g^+\cdot St$  & $Re_{p}^{+}=V_s^+ \cdot d_{p}^{+} / \nu^+$\\
\hline
0.2 & $0.227 \cdot 10^{-3}$ & $0.068$ & $~~~9.1$ & $0.0188$ & $0.00128$\\
1   & $1.133 \cdot 10^{-3}$ & $0.153$ & $~20.4$ & $0.0943$ & $0.01443$\\
5   & $5.660 \cdot 10^{-3}$ & $0.342$ & $~45.6$ & $0.4717$ & $0.16132$\\
25  & $28.32 \cdot 10^{-3}$ & $0.765$ & $102.0$ & $2.3584$ & $1.80418$\\
125 & $1.415 \cdot 10^{-1}$ & $1.71$  & $228$ & $11.792$ & $20.1643$\\
\end{tabular}
\vspace{0.3cm}
\caption{\label{part} Particle parameters for the $Re_{\tau}^l$ simulations.}
\end{center}
\end{small}
\end{minipage}
\end{table}
\vspace{-0.8cm}
%
%
\begin{table}[t]
\begin{minipage}[t]{15.0cm}
\begin{small}
\begin{center}
\begin{tabular}{c c c c c c}
$St^h=St |_{Re_{\tau}^h}$ & $\tau_{p}^h~(s)$ & $d_{p}^{+}$ & $d_{p}$ (${\mu}m$) & $V_{s}^{+}=g^+\cdot St$  & $Re_{p}^{+}=V_s^+ \cdot d_{p}^{+} / \nu^+$\\
\hline
1   & $0.283 \cdot 10^{-3}$ & $0.153$ & $10.2$ & $0.0118$ & $0.00275$\\
4   & $1.132 \cdot 10^{-3}$ & $0.306$ & $20.4$ & $0.0472$ & $0.01444$\\
5   & $1.415 \cdot 10^{-3}$ & $0.342$ & $22.8$ & $0.0590$ & $0.02018$\\
20  & $5.660 \cdot 10^{-3}$ & $0.684$ & $45.6$ & $0.2358$ & $0.16129$\\
25  & $7.075 \cdot 10^{-3}$ & $0.765$ & $51.0$ & $0.2948$ & $0.22552$\\
100  & $28.30 \cdot 10^{-3}$ & $1.530$ & $102.0$ & $1.1792$& $1.80418$\\
\end{tabular}
\vspace{0.3cm}
\caption{\label{part300} Particle parameters for the $Re_{\tau}^h$ simulations.}
\end{center}
\end{small}
\end{minipage}
\end{table}

%% file: figures.tex
%
%

\clearpage
\newpage

\begin{figure}
\centerline{\includegraphics[height=8.0cm]{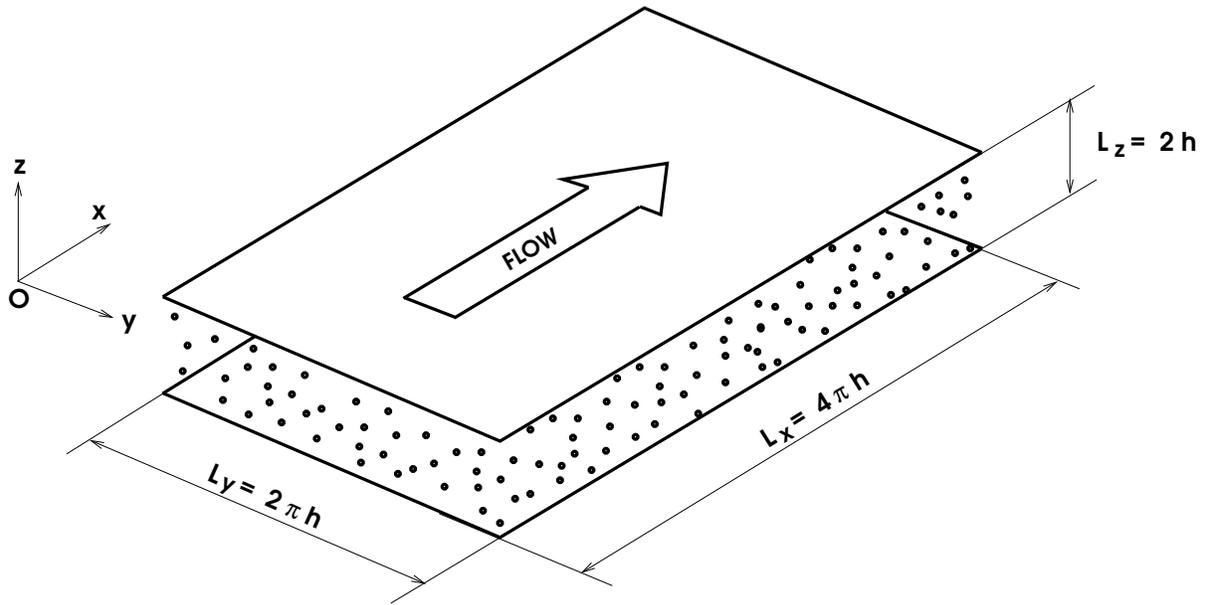}}
\caption{Particle-laden turbulent gas flow in a flat channel: computational domain.}
\label{f-chandomain}
\end{figure}

%
%

\clearpage
\newpage

\begin{figure}
\centerline{\includegraphics[height=8.5cm,angle=0]{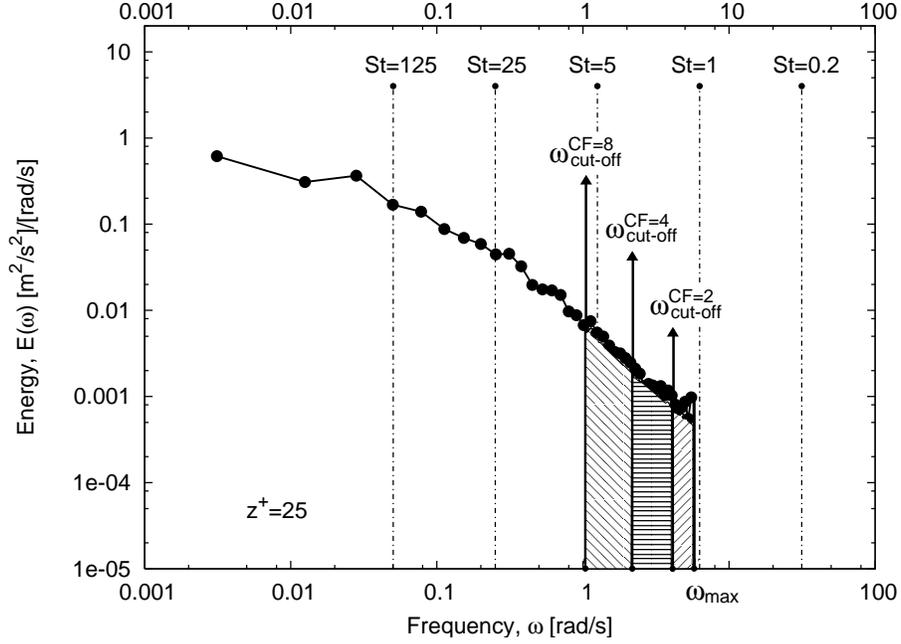}}
\vspace{0.2cm}
\caption{One-dimensional (streamwise) frequency spectrum for turbulent channel flow
at $Re_{\tau}=150$, computed at $z^+=25$ at $Re_{\tau}=150$. The different cut-off
frequencies, used to perform the {\em a-priori} tests, are indicated as
$\omega_{{\text{cut-off}}}^{{\text{CF=2}}}$,
$\omega_{{\text{cut-off}}}^{{\text{CF=4}}}$
and
$\omega_{{\text{cut-off}}}^{{\text{CF=8}}}$,
respectively. Areas filled with patterns below the energy
profile represent the relative
amount of energy removed by each cut-off.
}
\label{spectrum-les}
\end{figure}

%
%

\clearpage
\newpage

\begin{figure}
\centerline{\includegraphics[angle=270,width=5.5cm]{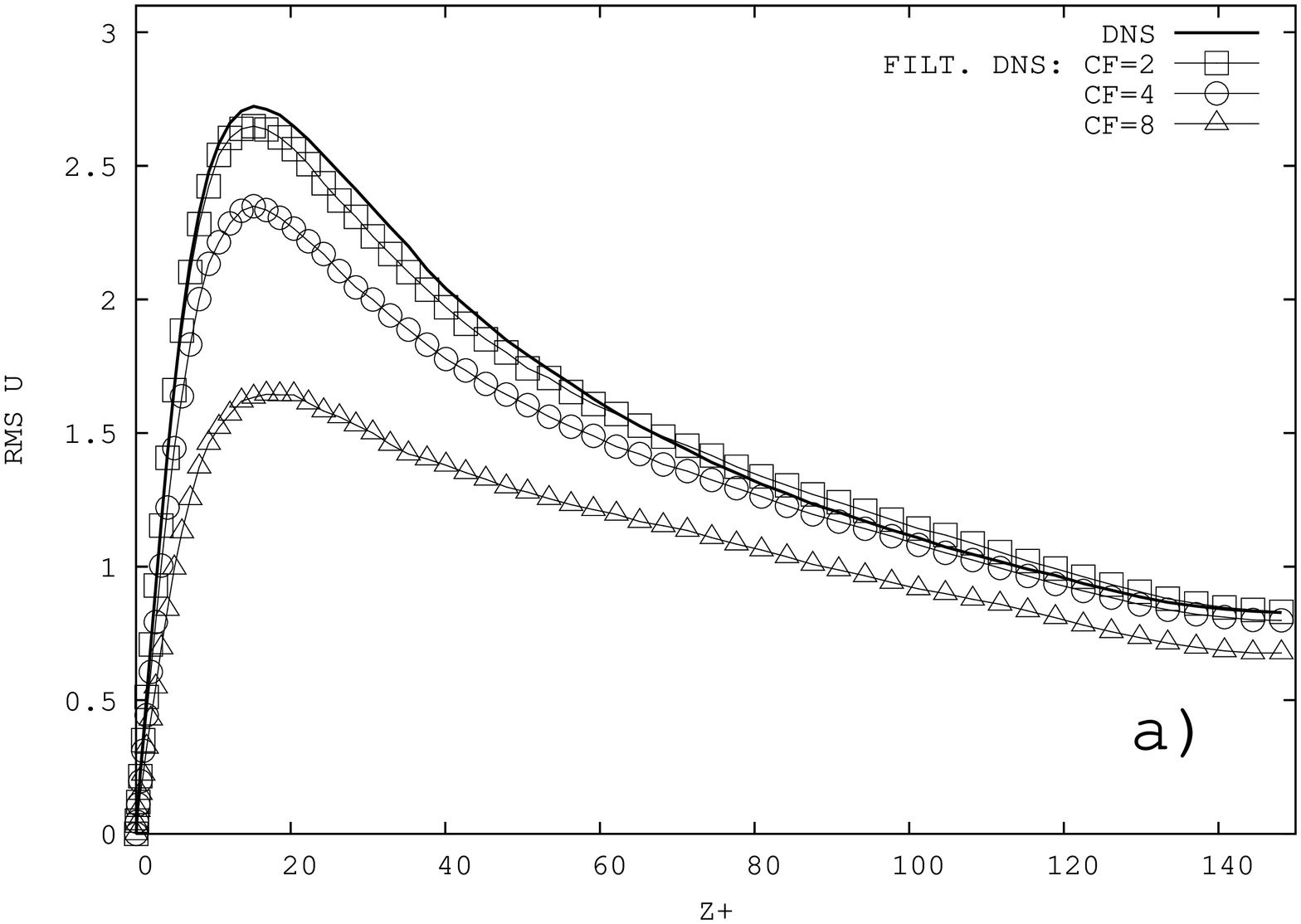}
\includegraphics[angle=270,width=5.5cm]{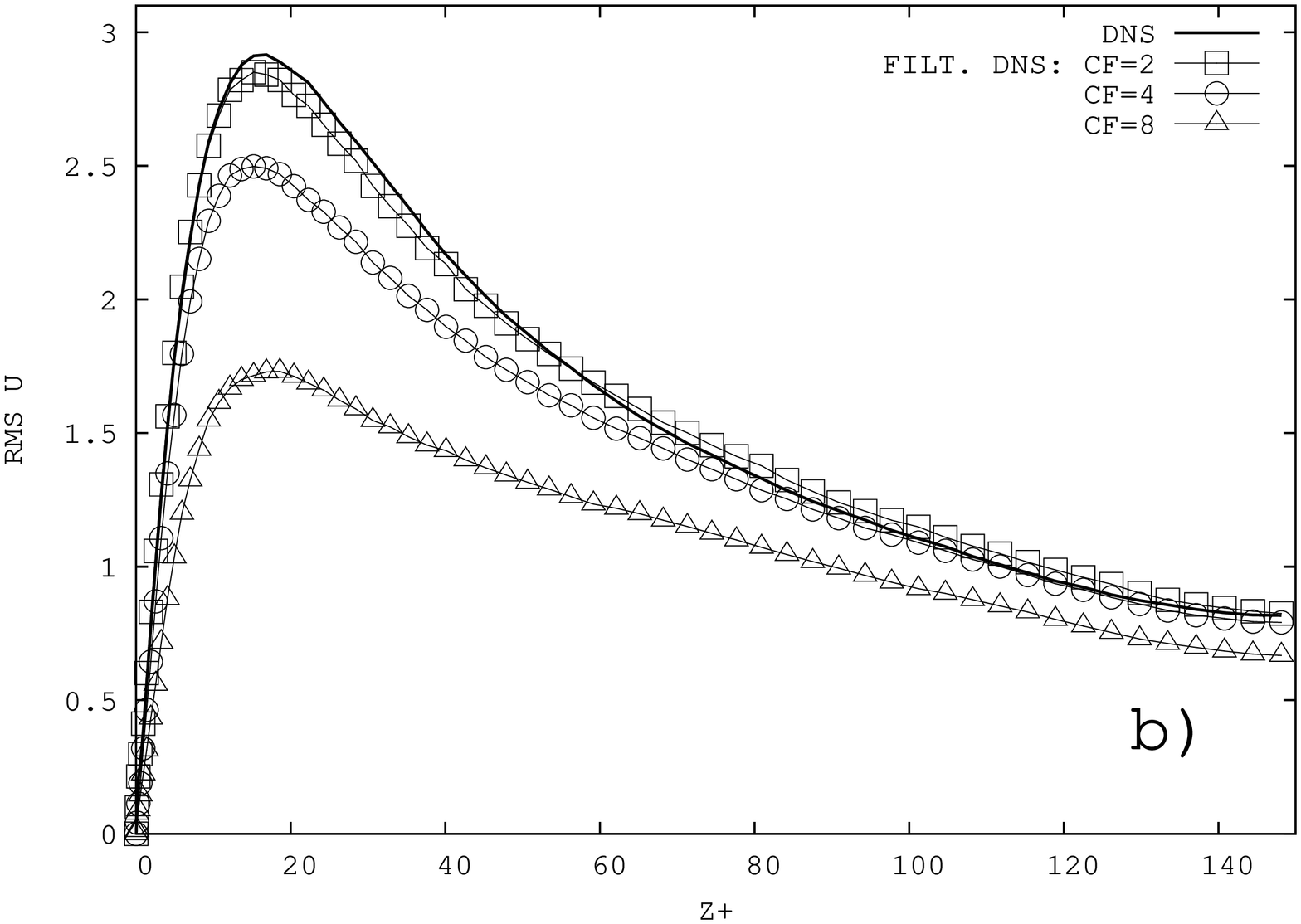}
\includegraphics[angle=270,width=5.5cm]{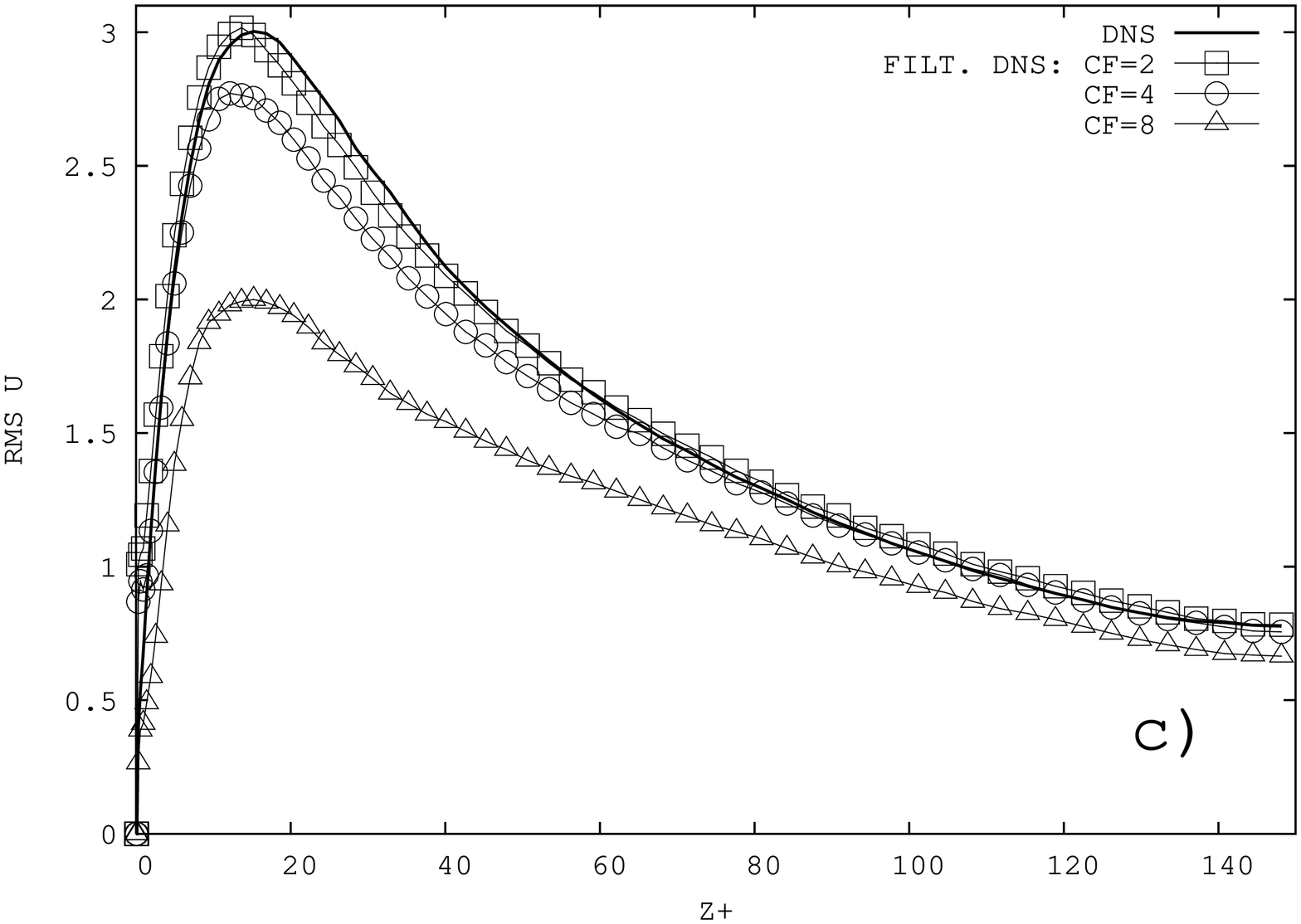}}
\centerline{\includegraphics[angle=270,width=5.5cm]{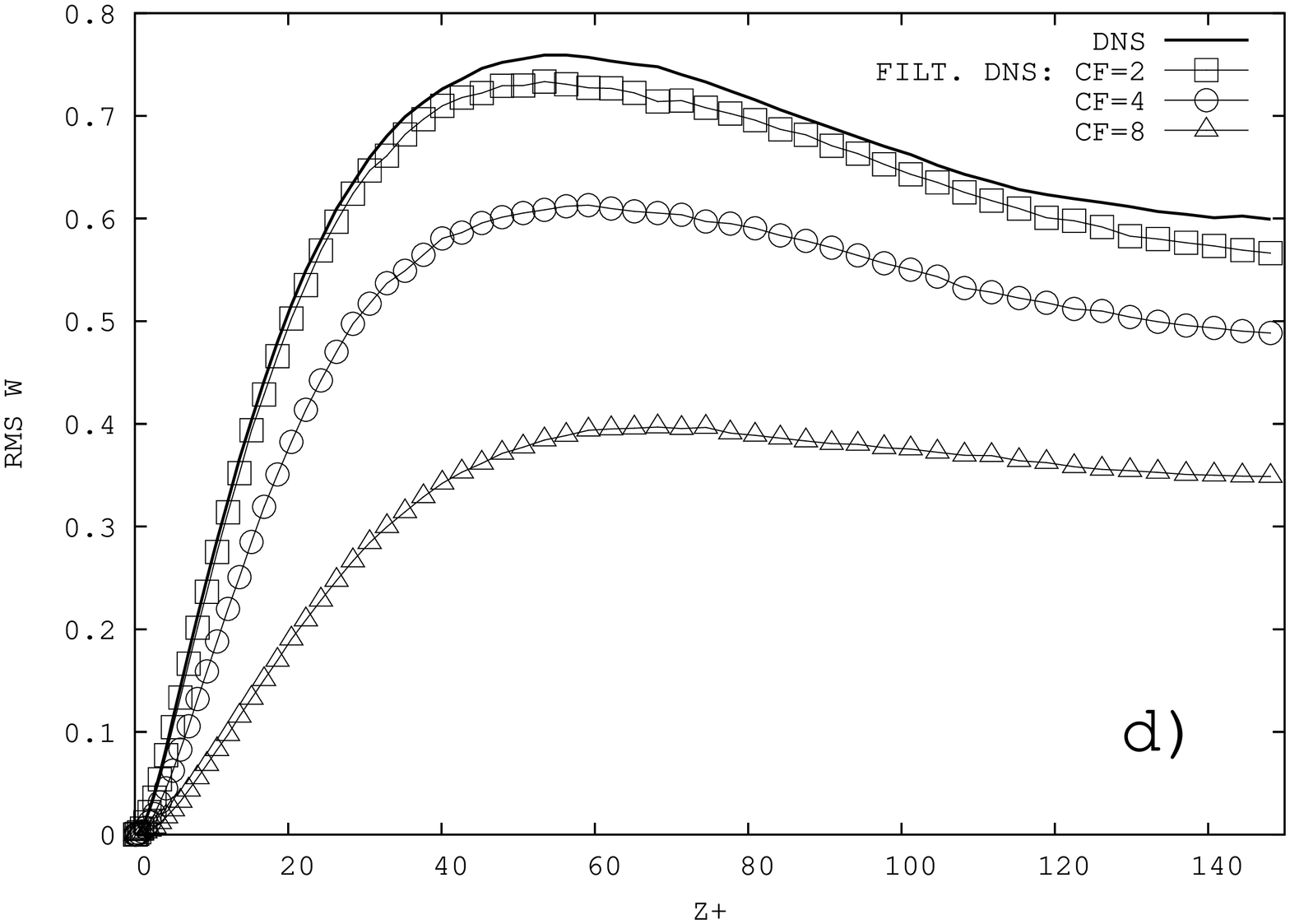}
\includegraphics[angle=270,width=5.5cm]{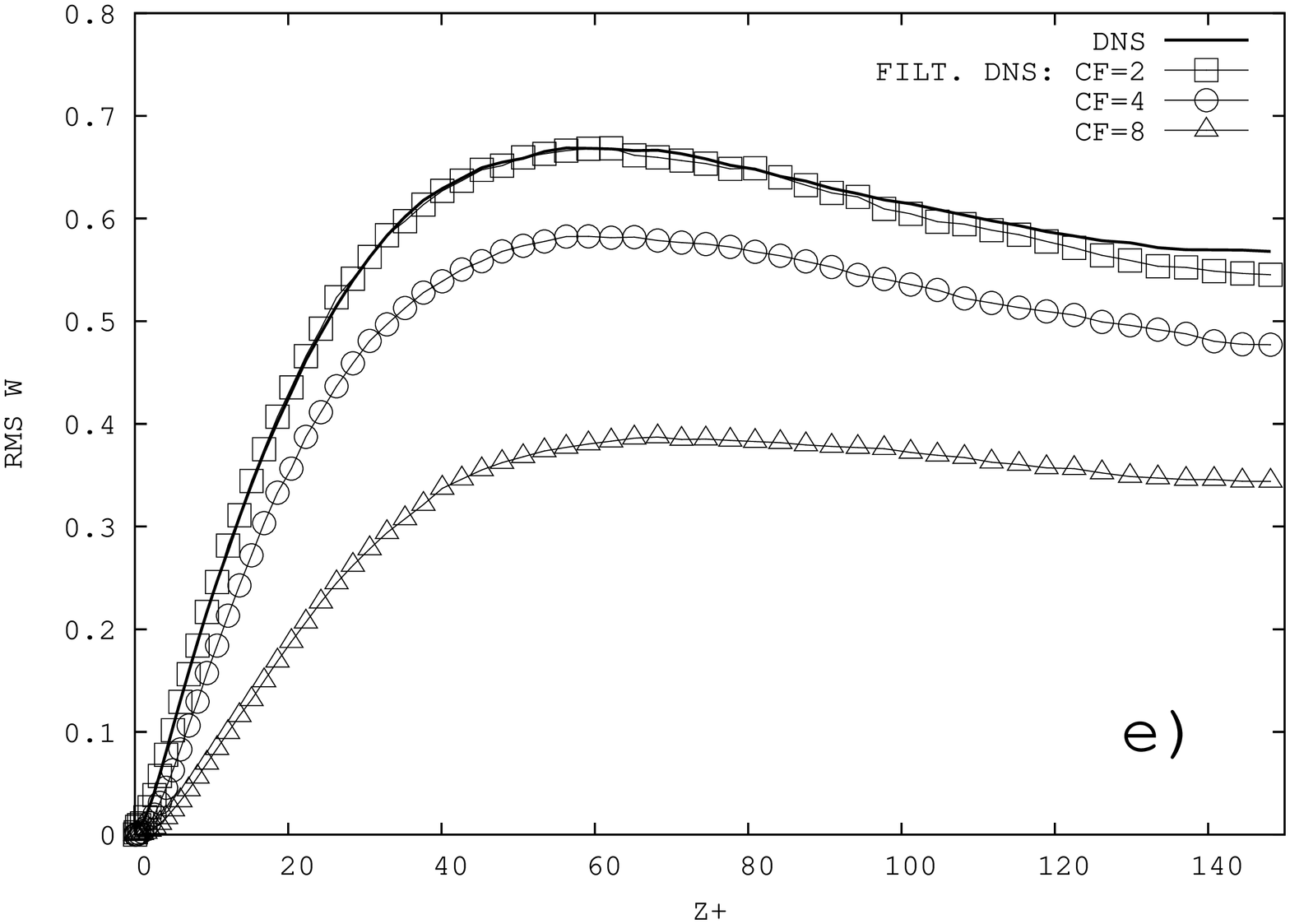}
\includegraphics[angle=270,width=5.5cm]{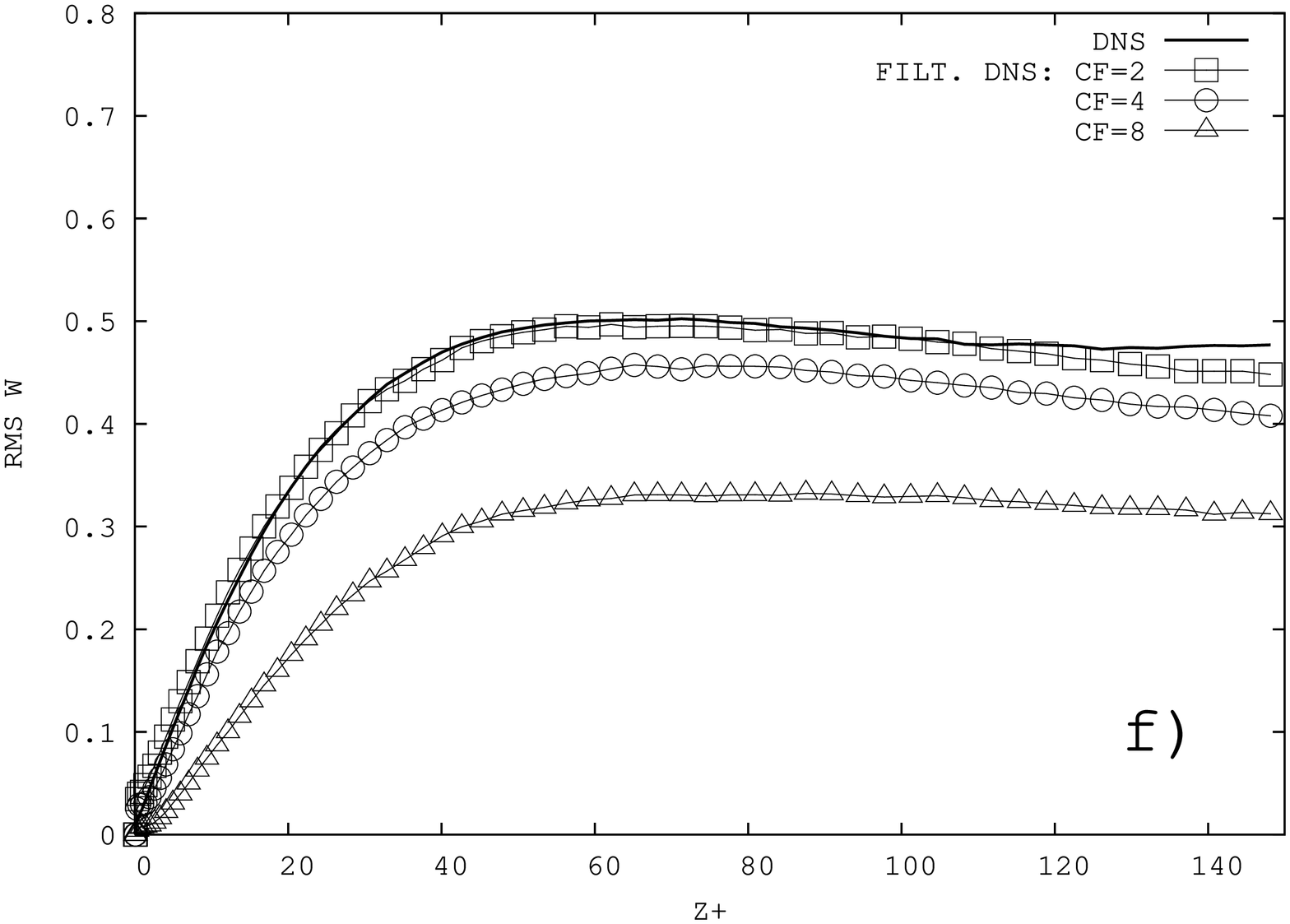}}
\vspace{0.4cm}
\caption{{\label{nfi}}
Particle rms velocity fluctuations for {\em a priori} simulations (with cut-off filter) without
SGS modeling in the
particle equation of motion: (a-c) streamwise rms component, (d-f) wall-normal rms
component.
Left-hand panels: $St=1$ particles, central panels: $St=5$ particles,
right-hand panels: $St=25$ particles.
CF indicates the LES grid coarsening factor with respect to the DNS grid:
CF=2 ($\square$), CF=4 ($\bigcirc$), CF=8 ($\triangle$).}
\end{figure}

%
%

\clearpage
\newpage

\begin{figure}
\centerline{\includegraphics[height=8.0cm,angle=270]{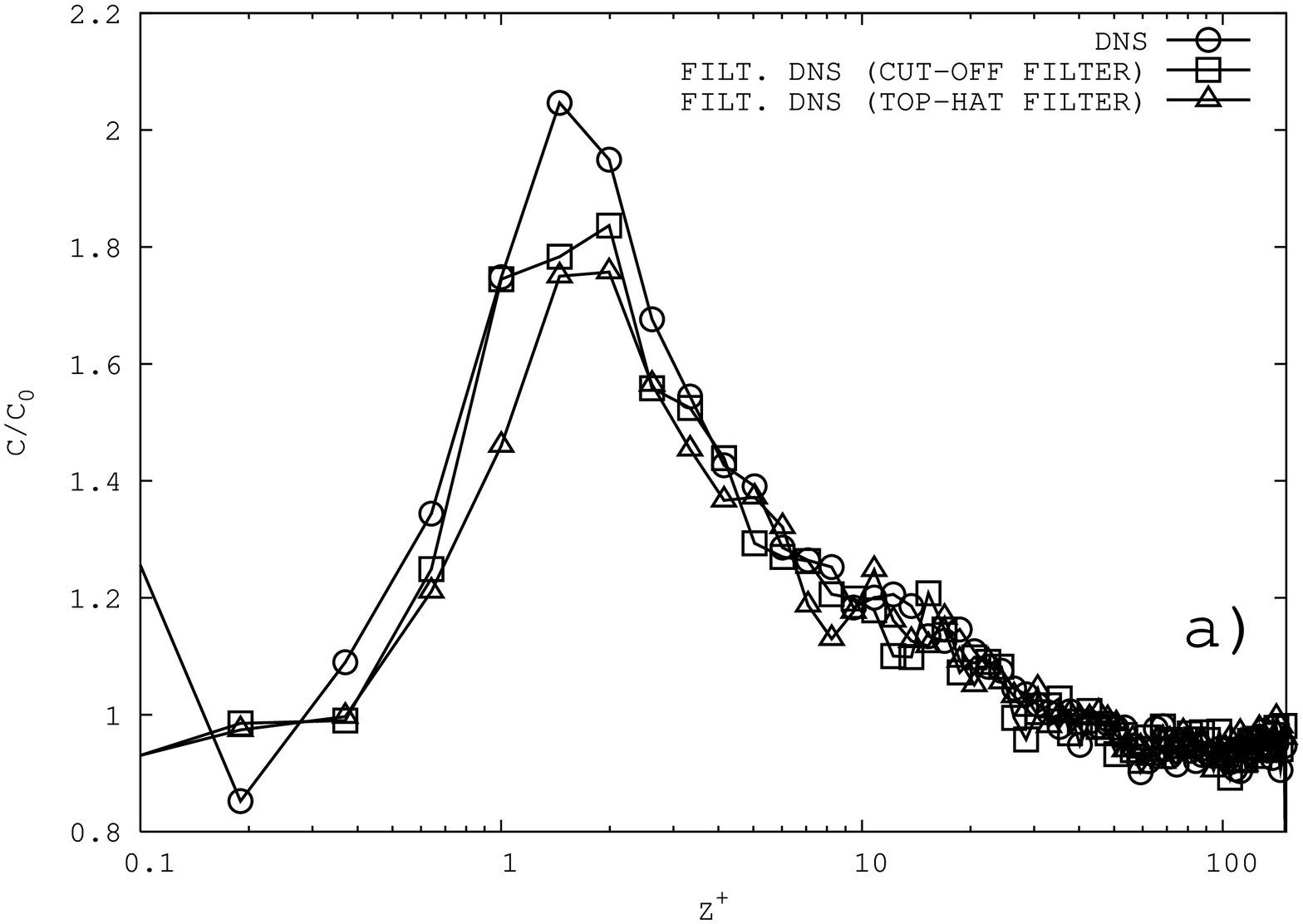}
\includegraphics[height=8.0cm,angle=270]{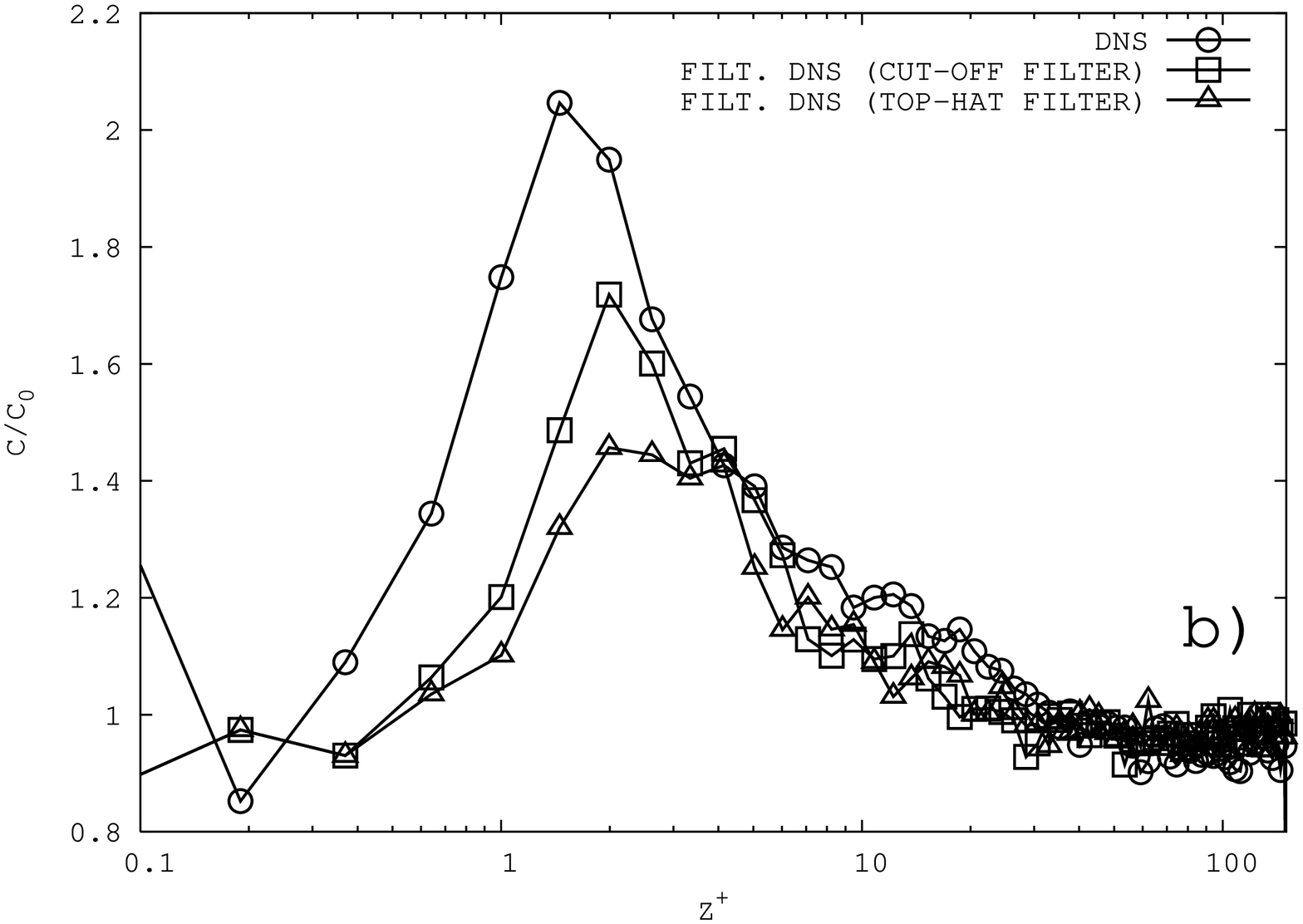}}
\centerline{\includegraphics[height=7.9cm,angle=270]{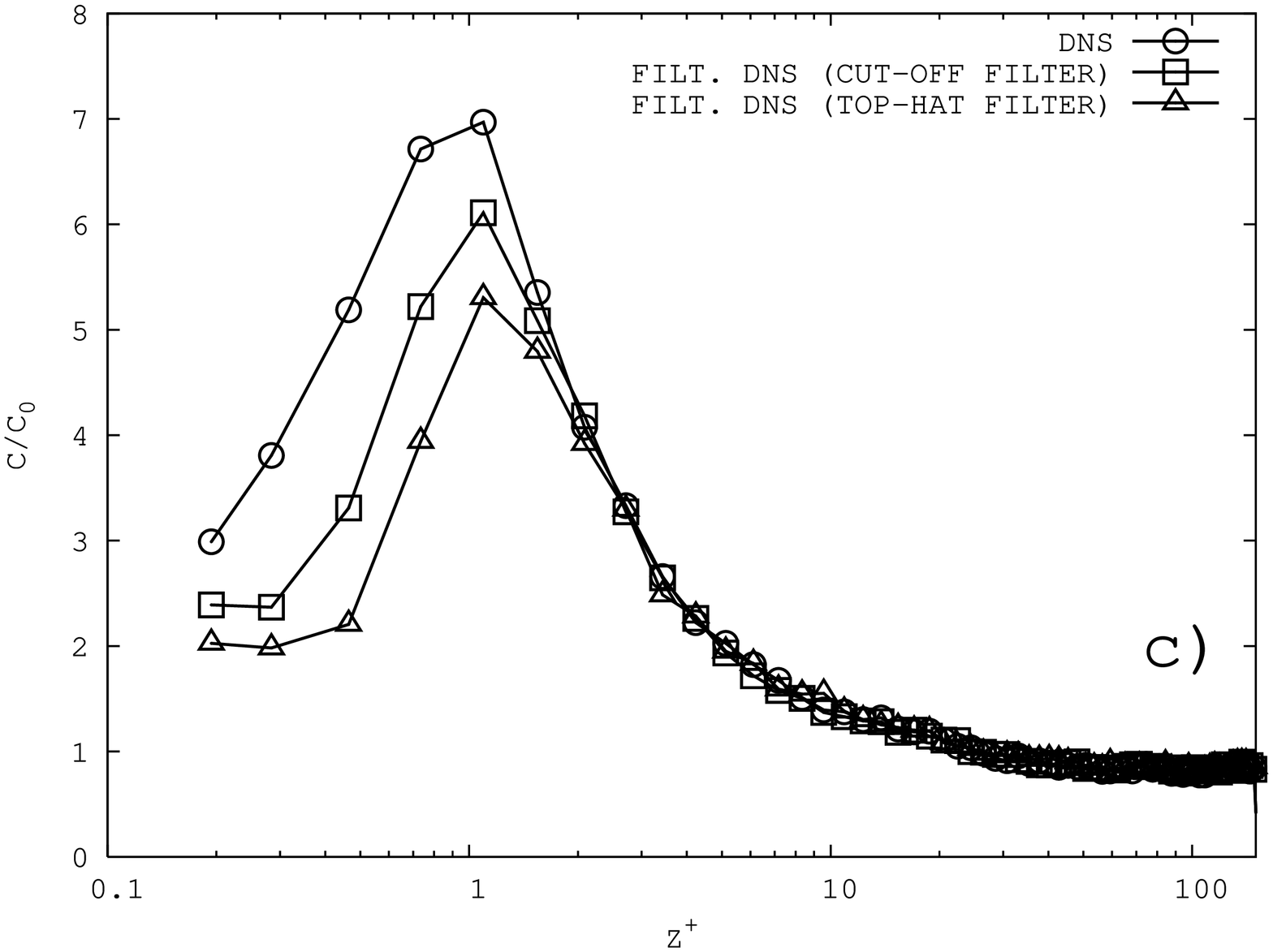}
\hspace{0.2cm}
\includegraphics[height=7.9cm,angle=270]{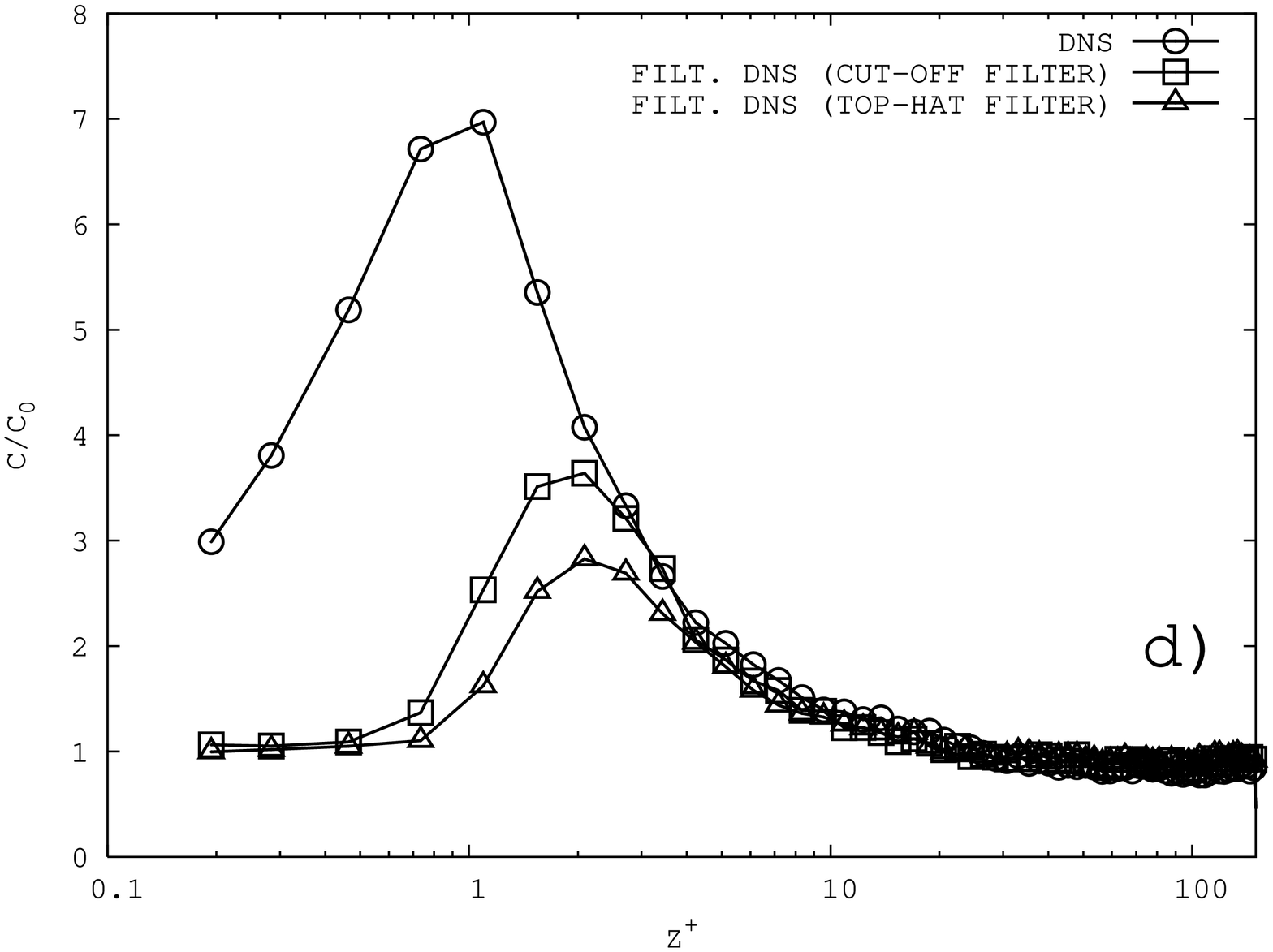}}
\centerline{\includegraphics[height=8.0cm,angle=270]{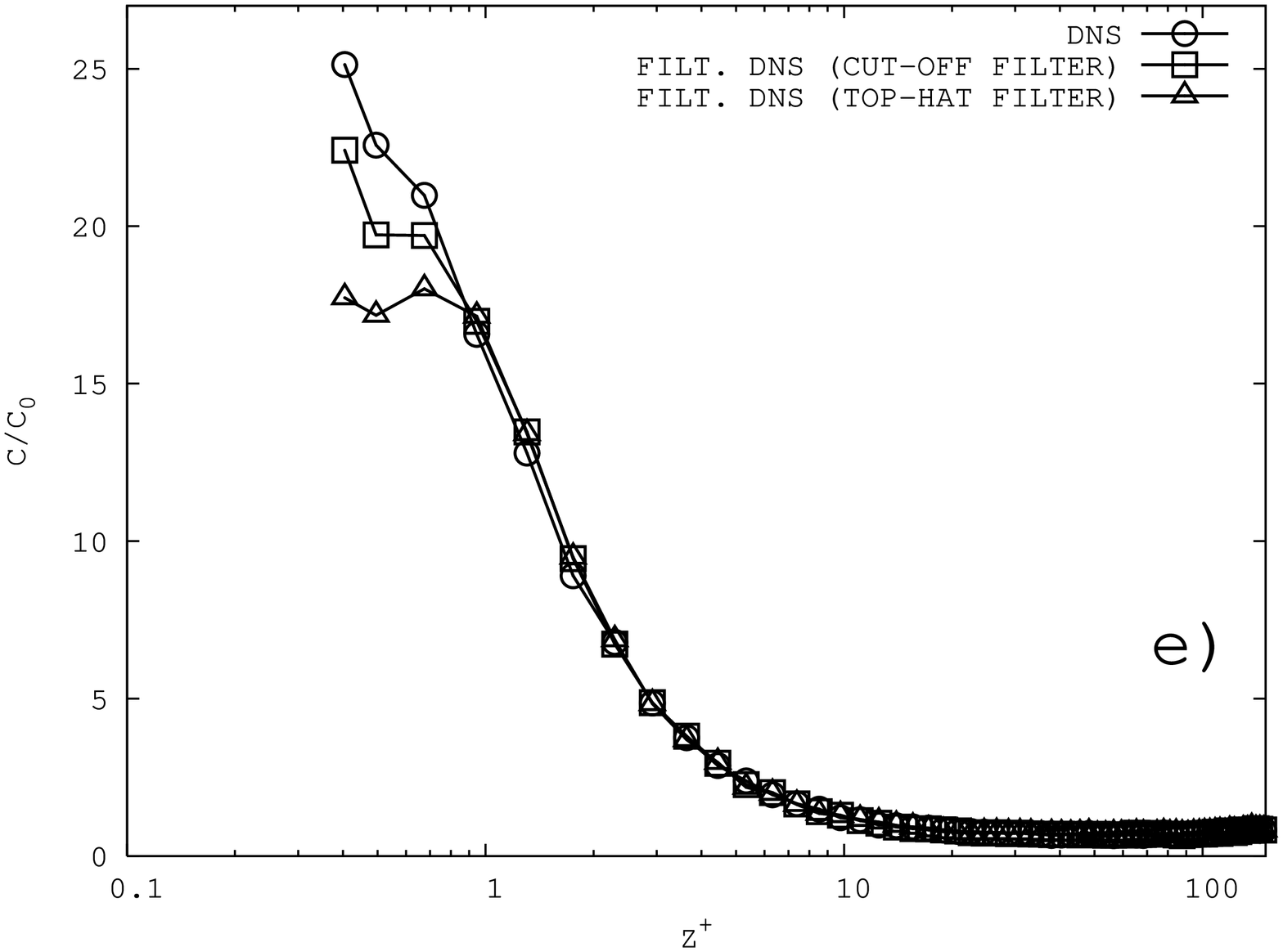}
\includegraphics[height=8.0cm,angle=270]{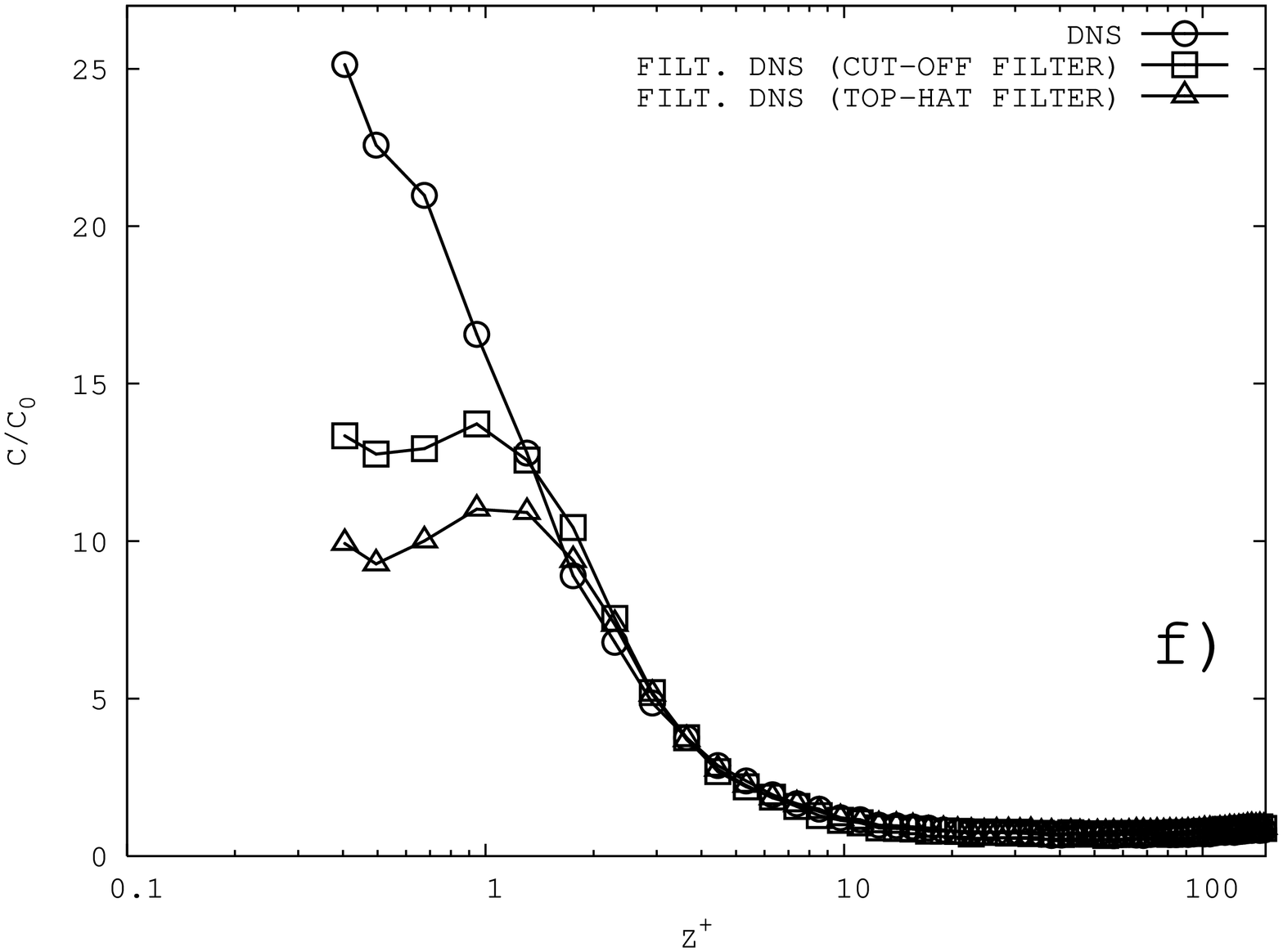}}
\vspace{0.2cm}
\caption{
Particle concentration in {\em a-priori} tests without SGS modeling
in the particle equation of motion: (a-b) $St=1$ particles, (c-d) $St=5$ particles,
(e-f) $St=25$ particles. DNS ($\bigcirc$), {\em a-priori} LES with
cut-off filter ($\square$), {\em a-priori} LES with top-hat filter ($\triangle$).
Left-hand panels: tests on the fine $64 \times 64 \times 65$ grid (CF=2), right-hand panels:
tests on the coarse $32 \times 32 \times 65$ grid (CF=4).
}
\label{a-priori-filtering}
\end{figure}

%
%

\clearpage
\newpage

\begin{figure}
\centerline{\includegraphics[height=12.0cm,angle=270]{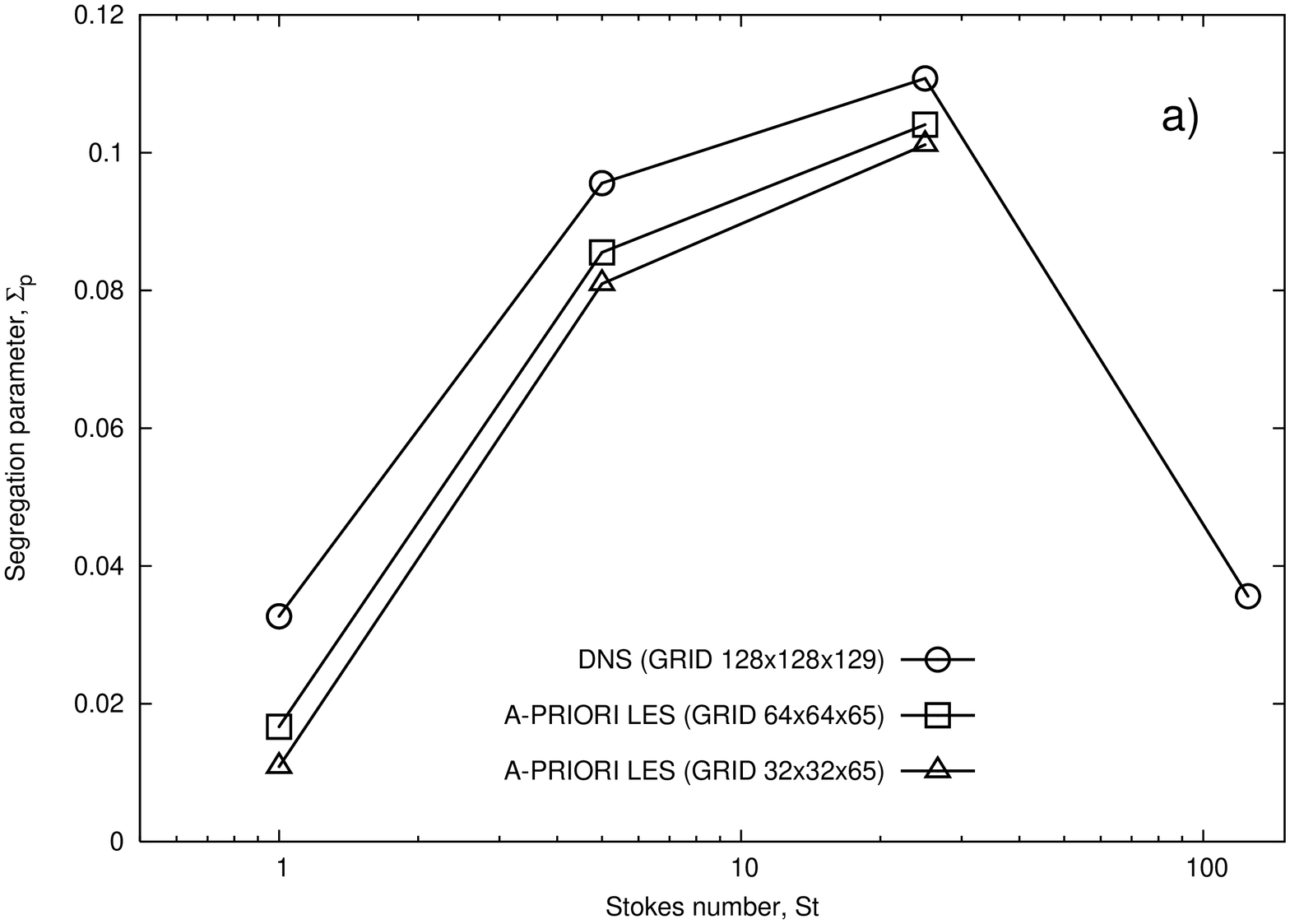}}
\centerline{\includegraphics[height=12.0cm,angle=270]{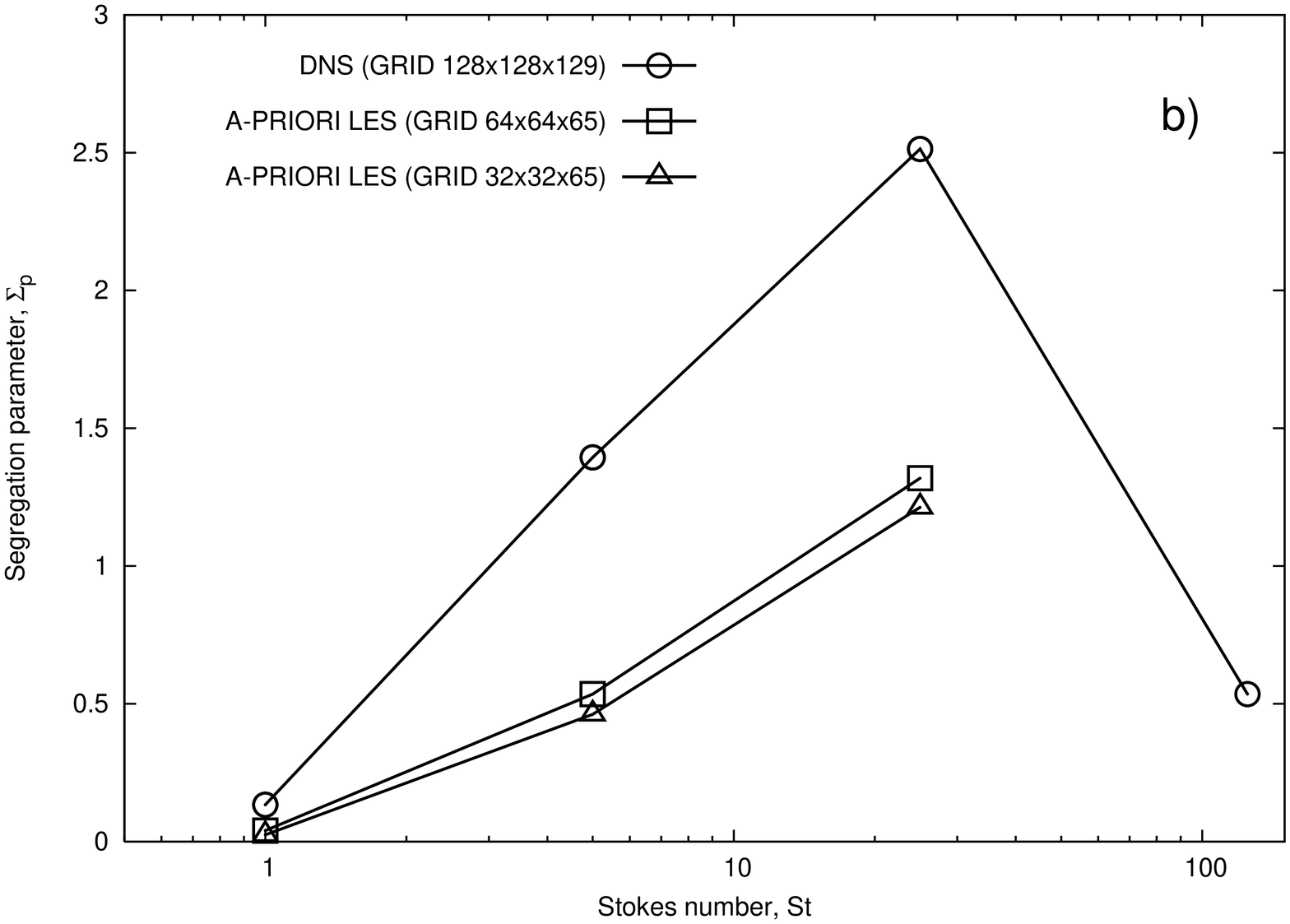}}
\vspace{0.2cm}
\caption{
Particle segregation, $\Sigma_p$, versus particle Stokes number, $St$,
in turbulent channel flow: comparison between DNS ($\bigcirc$),
{\em a-priori} LES on the fine $64 \times 64 \times 65$ grid ($\square$)
and {\em a-priori} LES on the coarse $32 \times 32 \times 65$ grid ($\triangle$).
{\em A-priori} results are relative to the cut-off filter.
Panels: (a) channel centerline ($145 \le z^+ \le 150$), (b) near-wall region
($0 \le z^+ \le 5$).
}
\label{poisson-dns-a-priori-les}
\end{figure}

%
%

\clearpage
\newpage

\begin{figure}
\centerline{\includegraphics[height=11.0cm,angle=270]{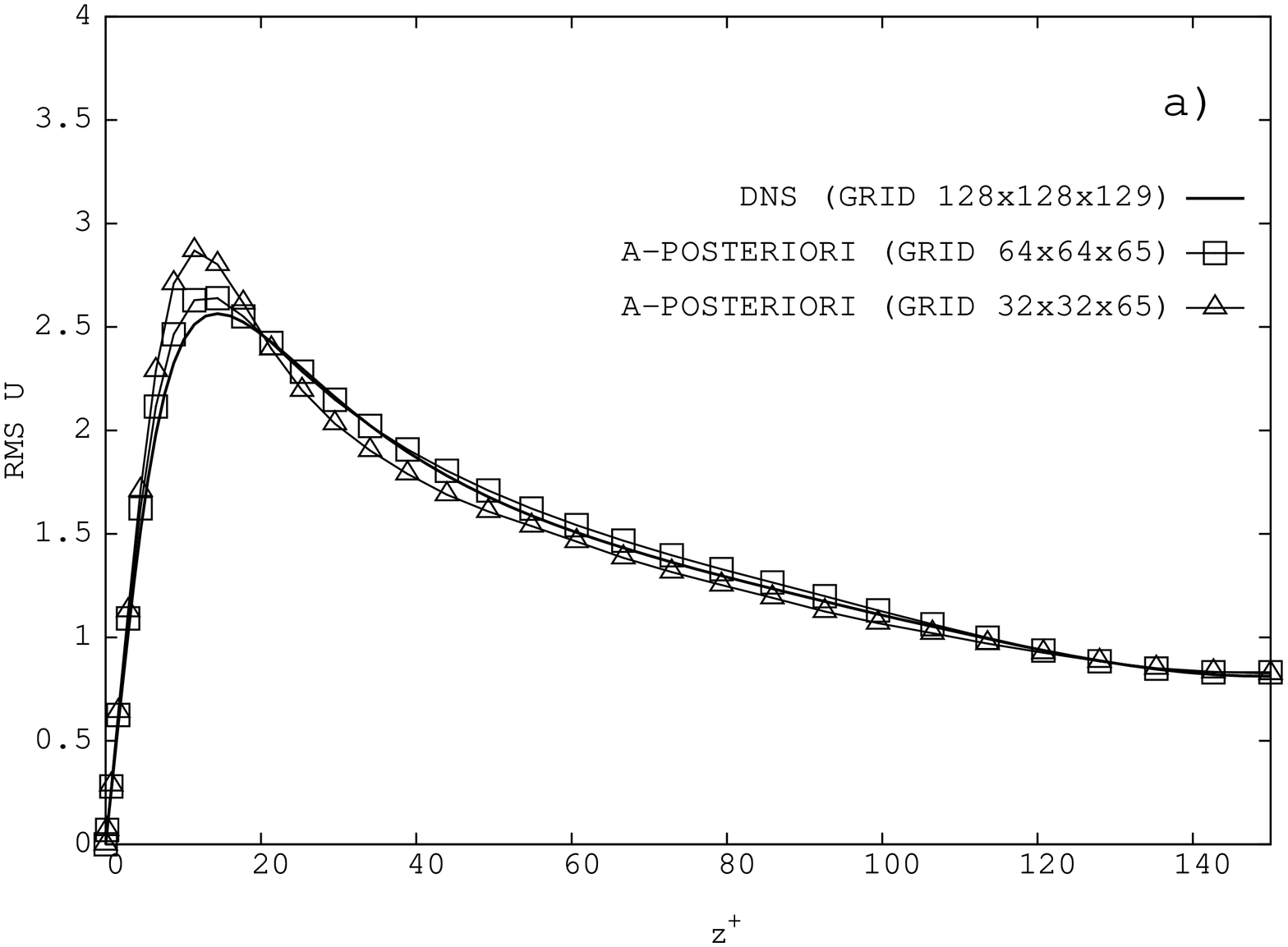}}
\centerline{\includegraphics[height=11.0cm,angle=270]{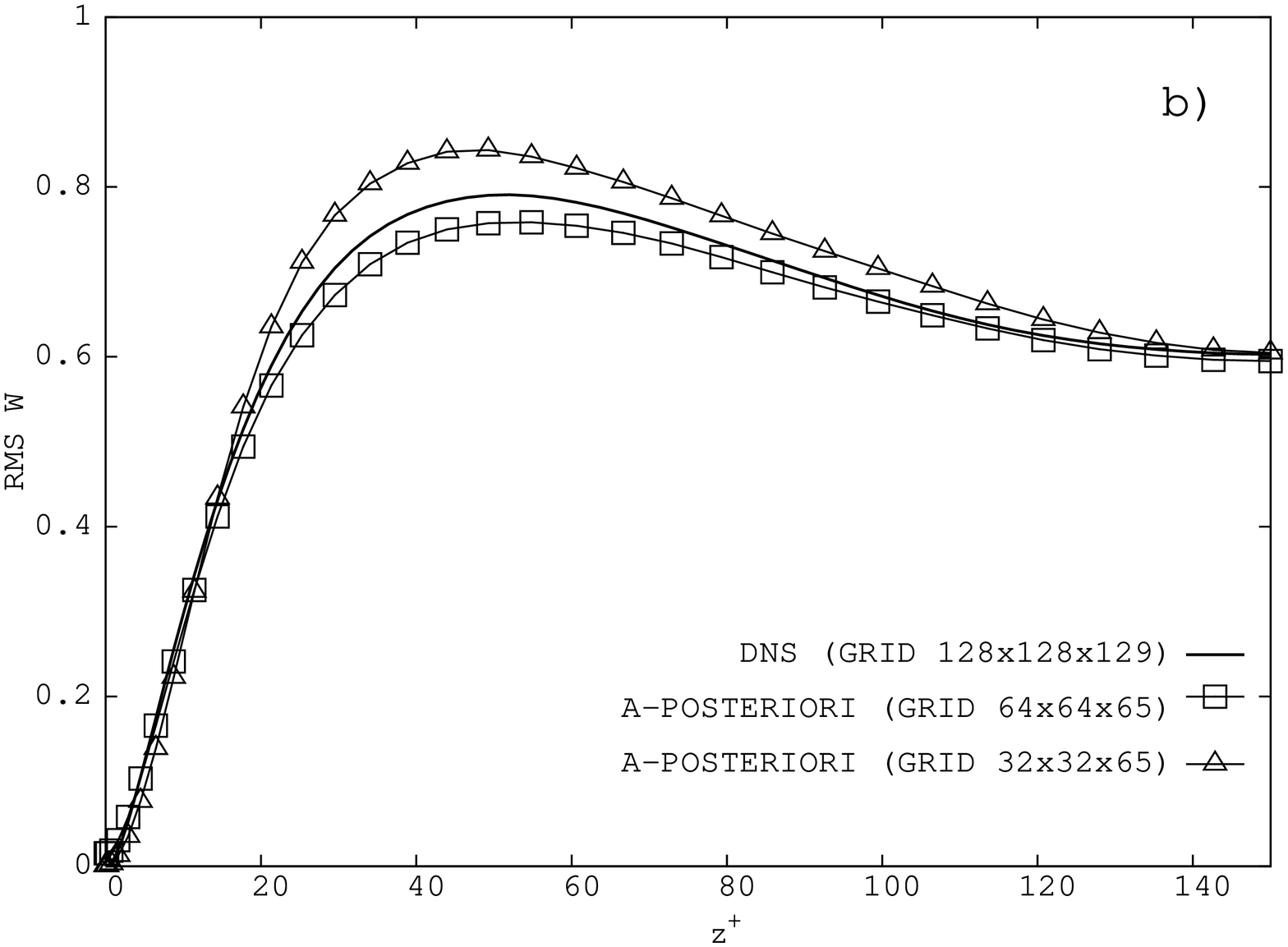}}
\vspace{0.5cm}
\caption{
Fluid rms velocity fluctuations: comparison between DNS (solid line), {\em a-posteriori}
LES on the fine $64 \times 64 \times 65$ grid ($\square$)
and {\em a-posteriori} LES on the coarse $32 \times 32 \times 65$ grid ($\triangle$).
Panels: (a) streamwise rms component, (b) wall-normal rms
component.
}
\label{rms-dns-les}
\end{figure}

%
%

\clearpage
\newpage

\begin{figure}
\centerline{\includegraphics[height=5.5cm,angle=270]{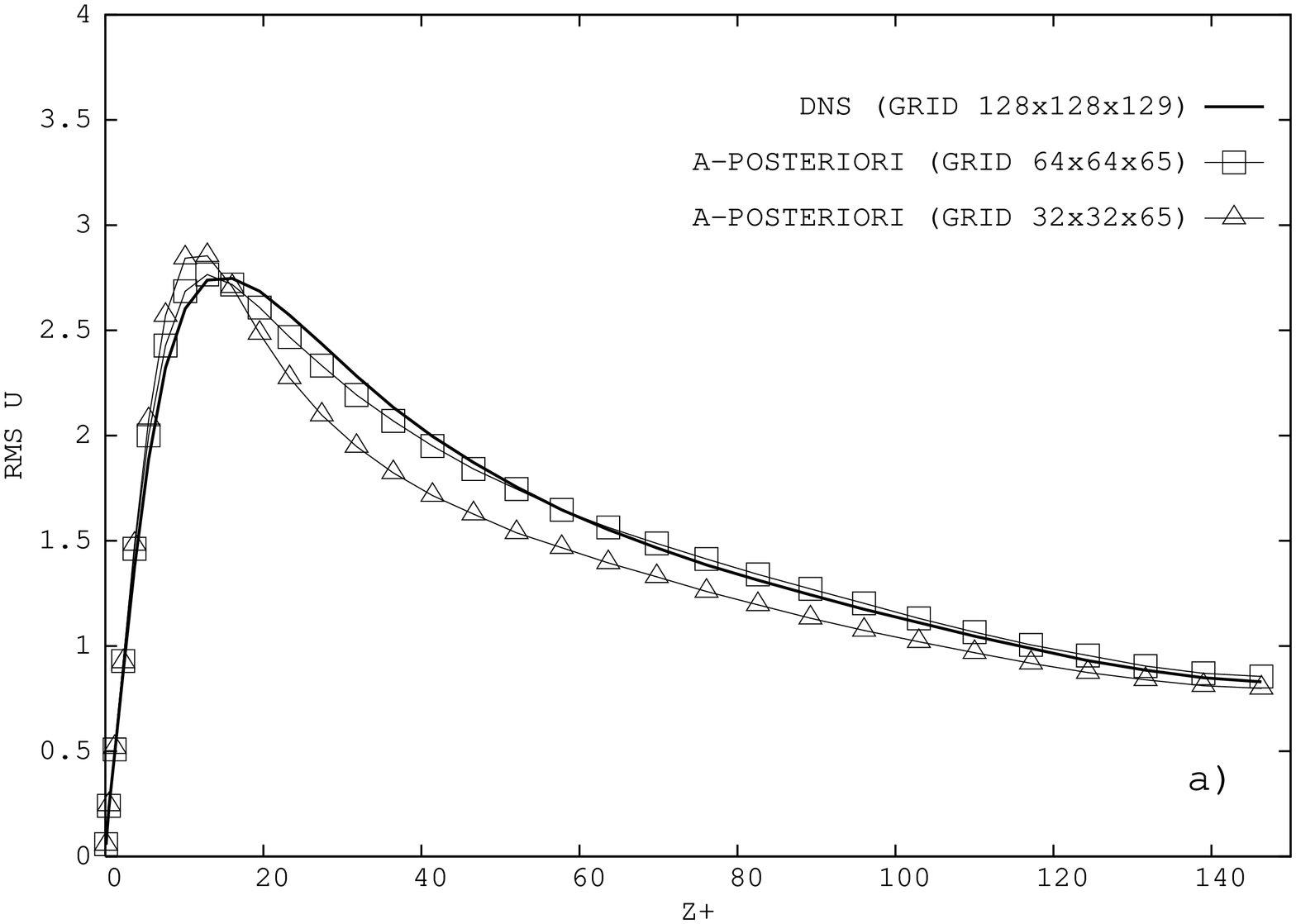}
\includegraphics[height=5.5cm,angle=270]{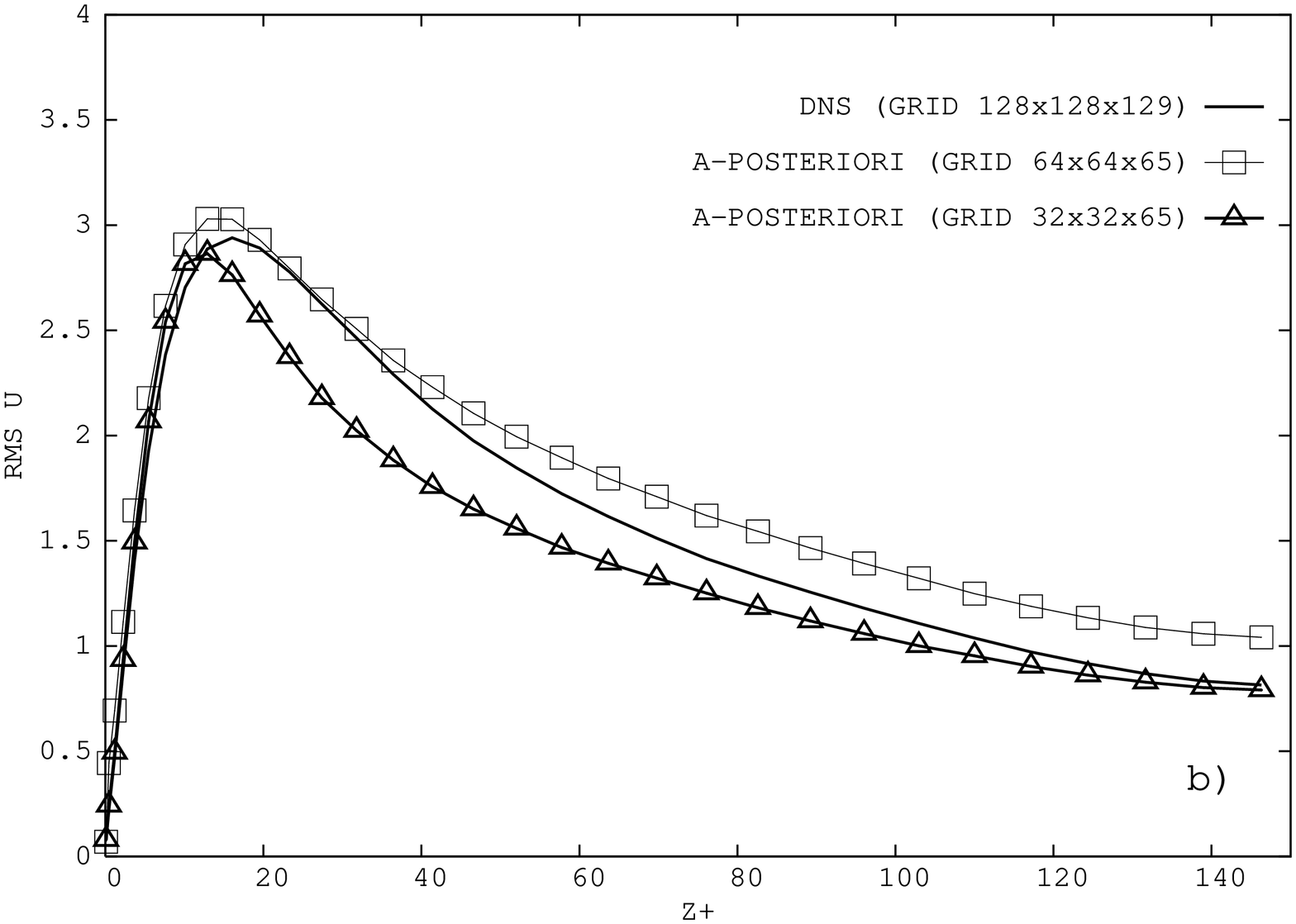}
\includegraphics[height=5.5cm,angle=270]{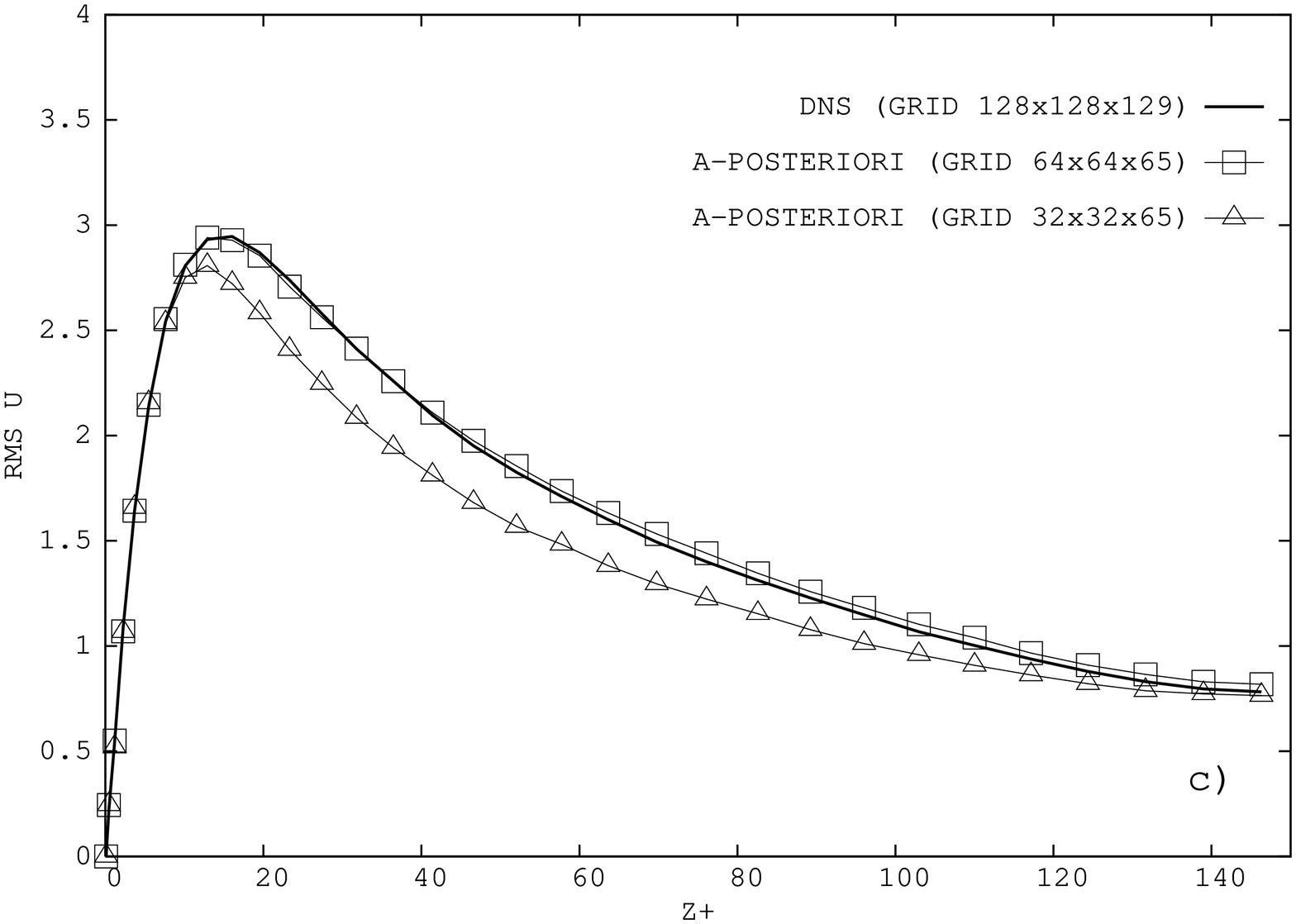}}
\centerline{\includegraphics[height=5.5cm,angle=270]{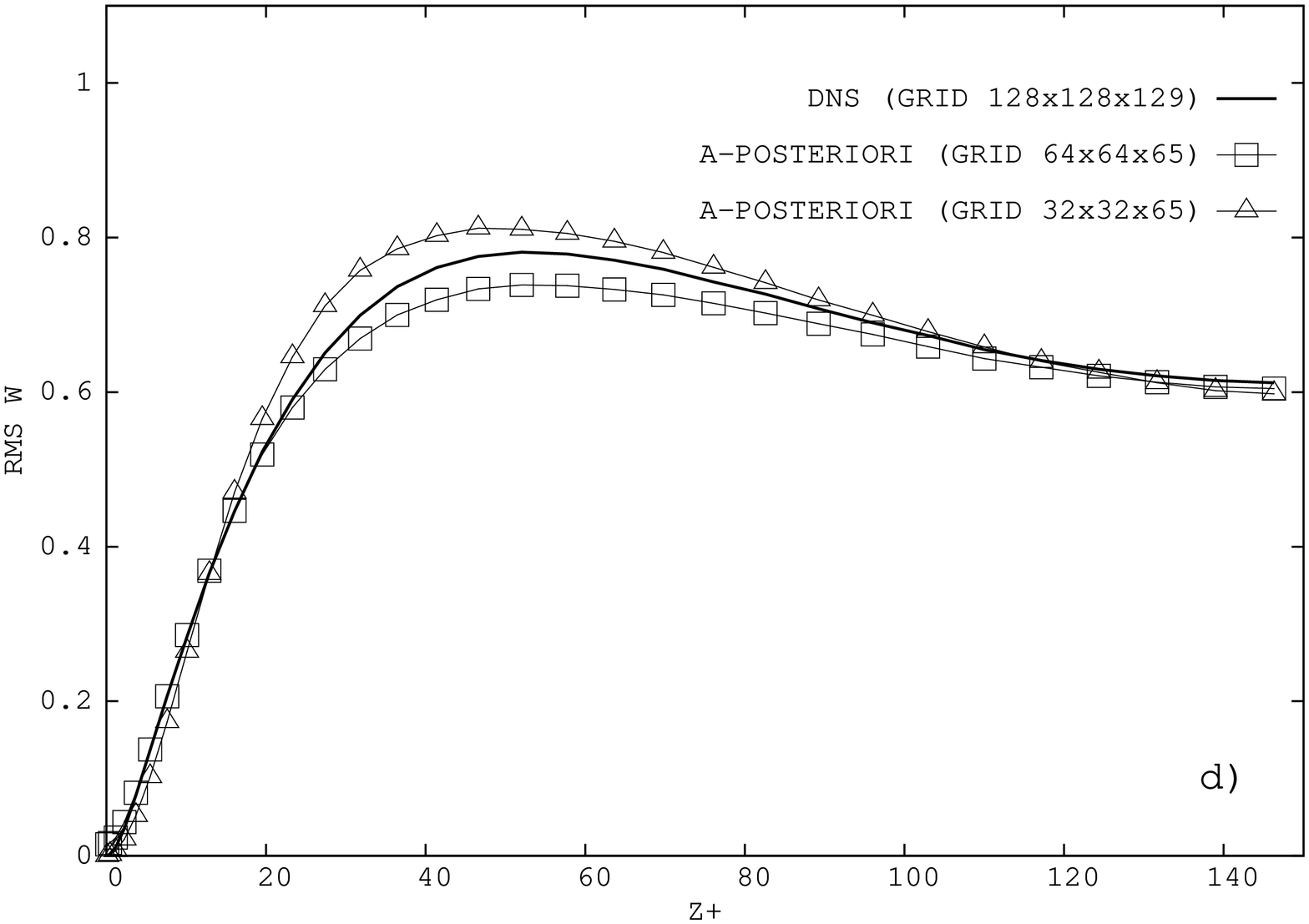}
\includegraphics[height=5.5cm,angle=270]{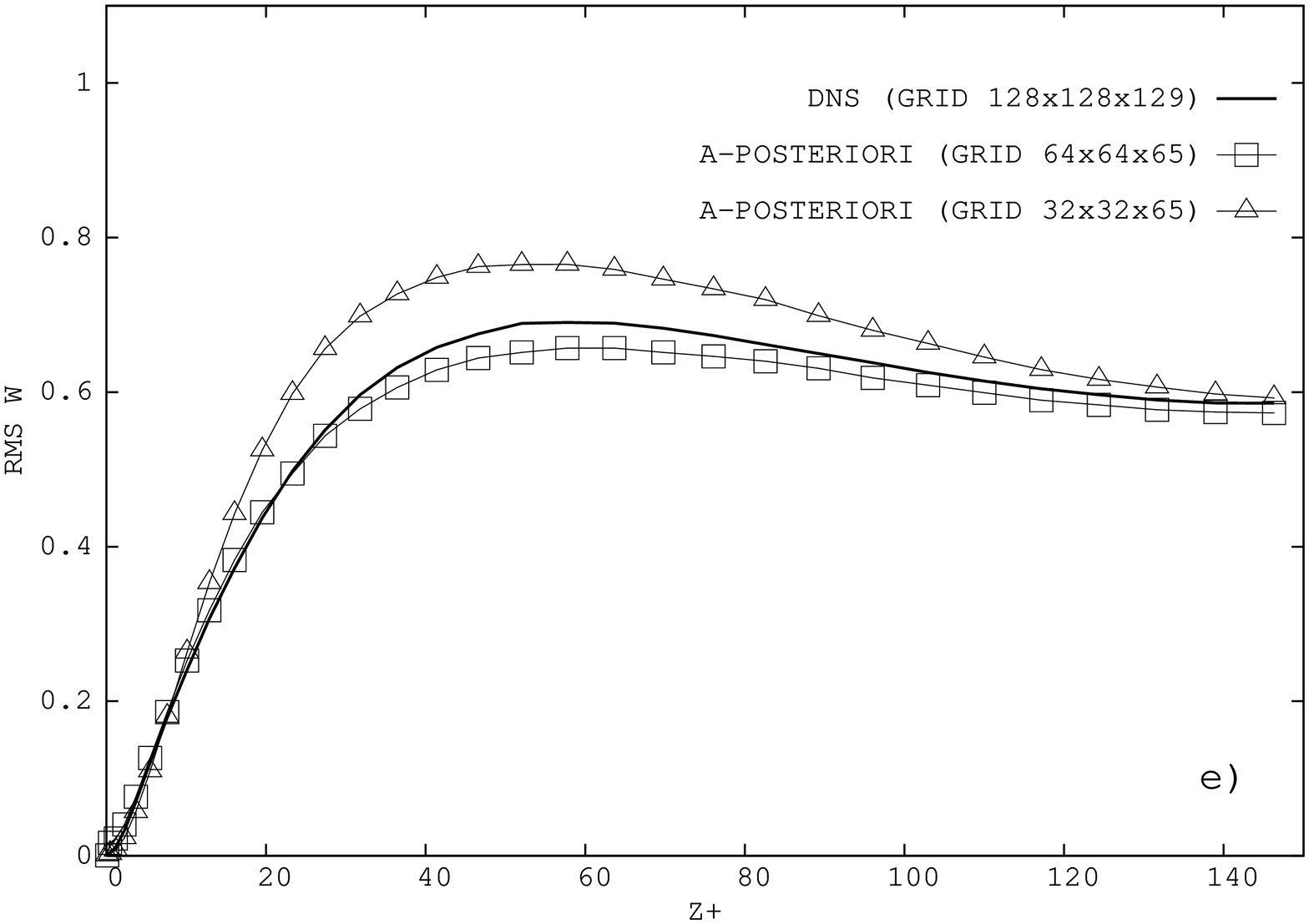}
\includegraphics[height=5.5cm,angle=270]{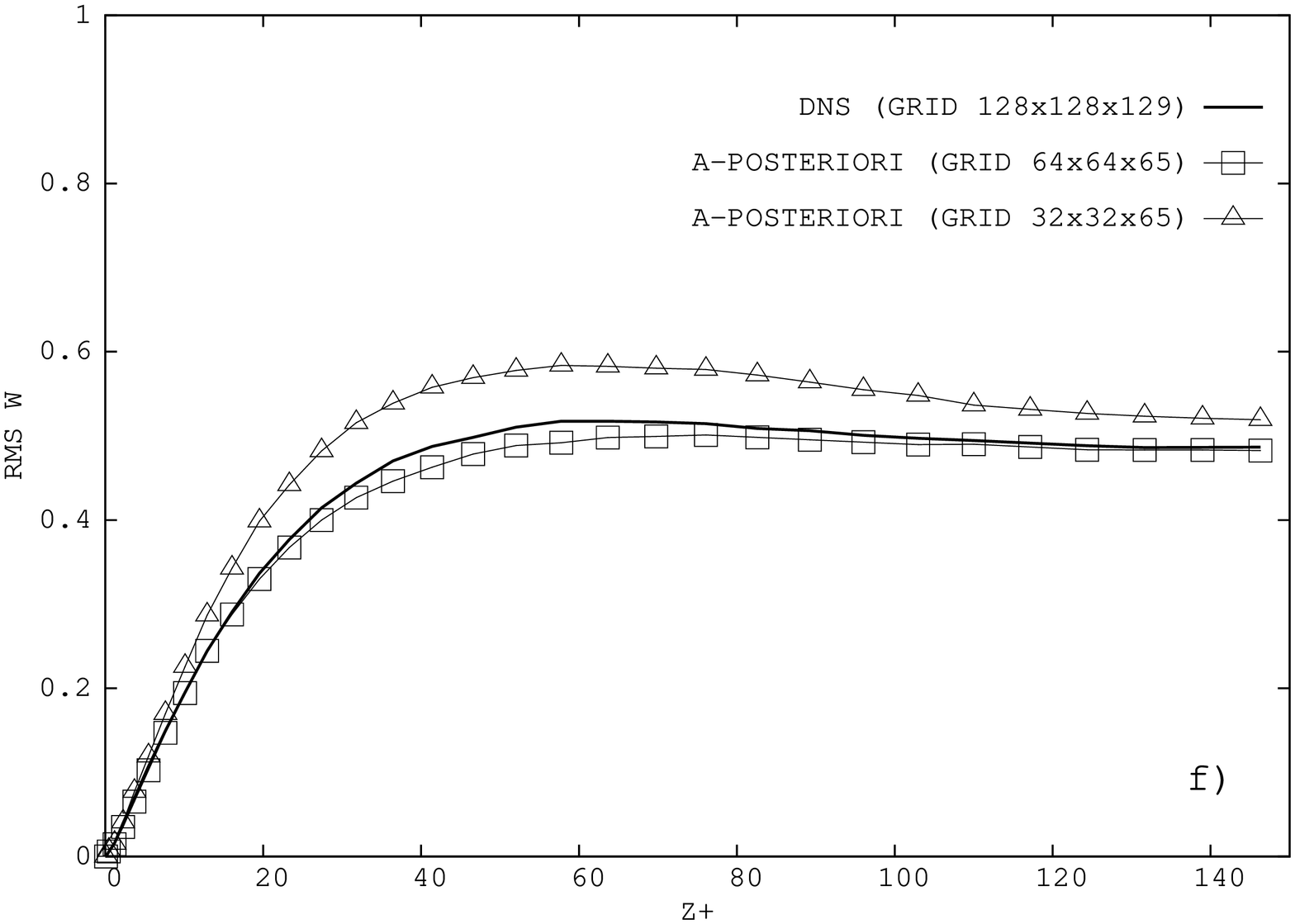}}
\vspace{0.5cm}
\caption{
Particle rms velocity fluctuations: comparison between DNS (solid line), {\em a-posteriori}
LES on the fine $64 \times 64 \times 65$ grid ($\square$)
and {\em a-posteriori} LES on the coarse $32 \times 32 \times 65$ grid ($\triangle$):
(a-c) streamwise rms component, (d-f) wall-normal rms
component.
Left-hand panels: $St=1$ particles, central panels: $St=5$ particles,
right-hand panels: $St=25$ particles.
}
\label{rmspart-dns-les}
\end{figure}

%
%

\clearpage
\newpage

\begin{figure}
\centerline{\includegraphics[height=9.0cm,angle=270]{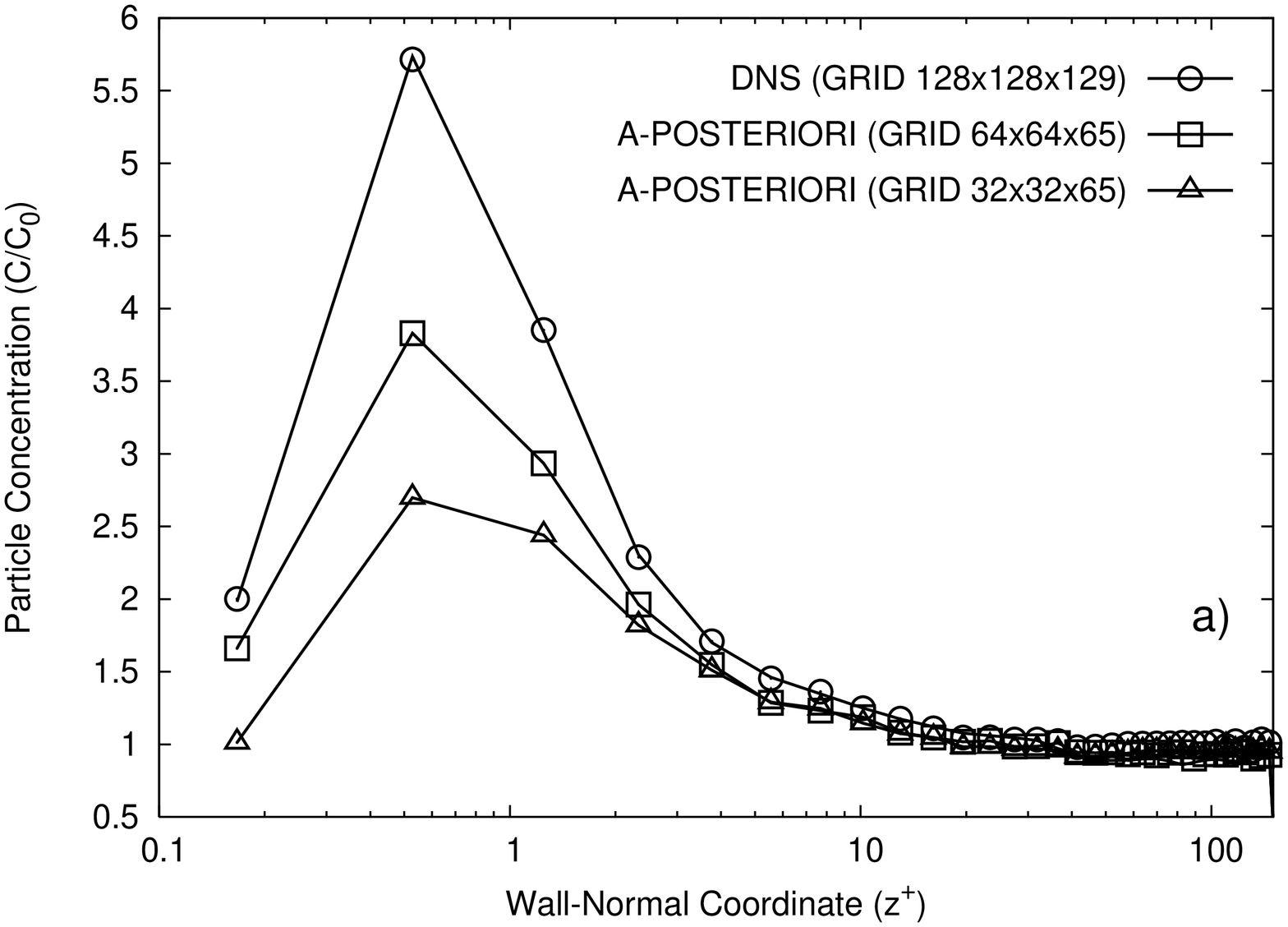}}
\vspace{-0.1cm}
\centerline{\includegraphics[height=9.0cm,angle=270]{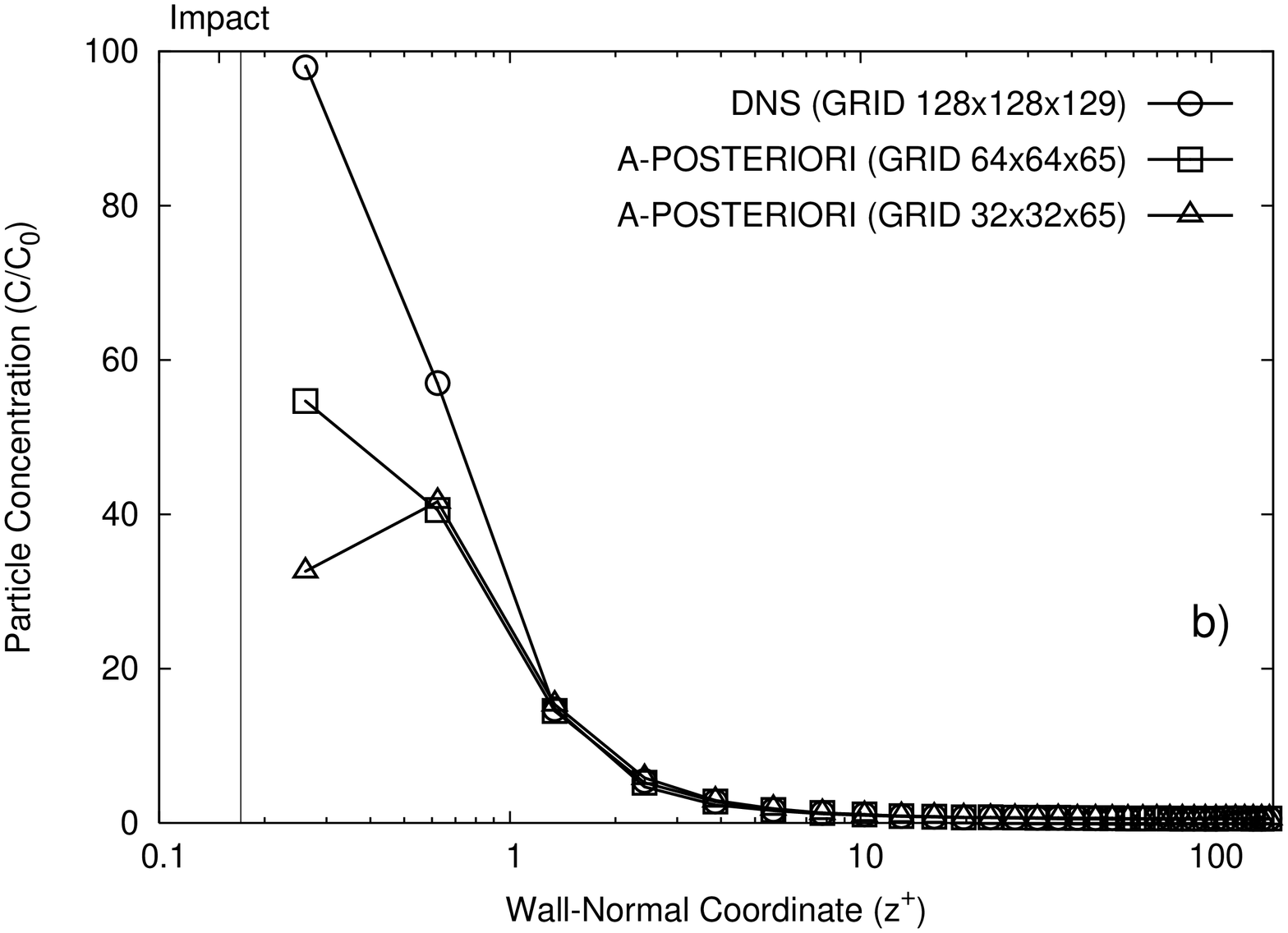}}
\vspace{-0.1cm}
\centerline{\includegraphics[height=9.0cm,angle=270]{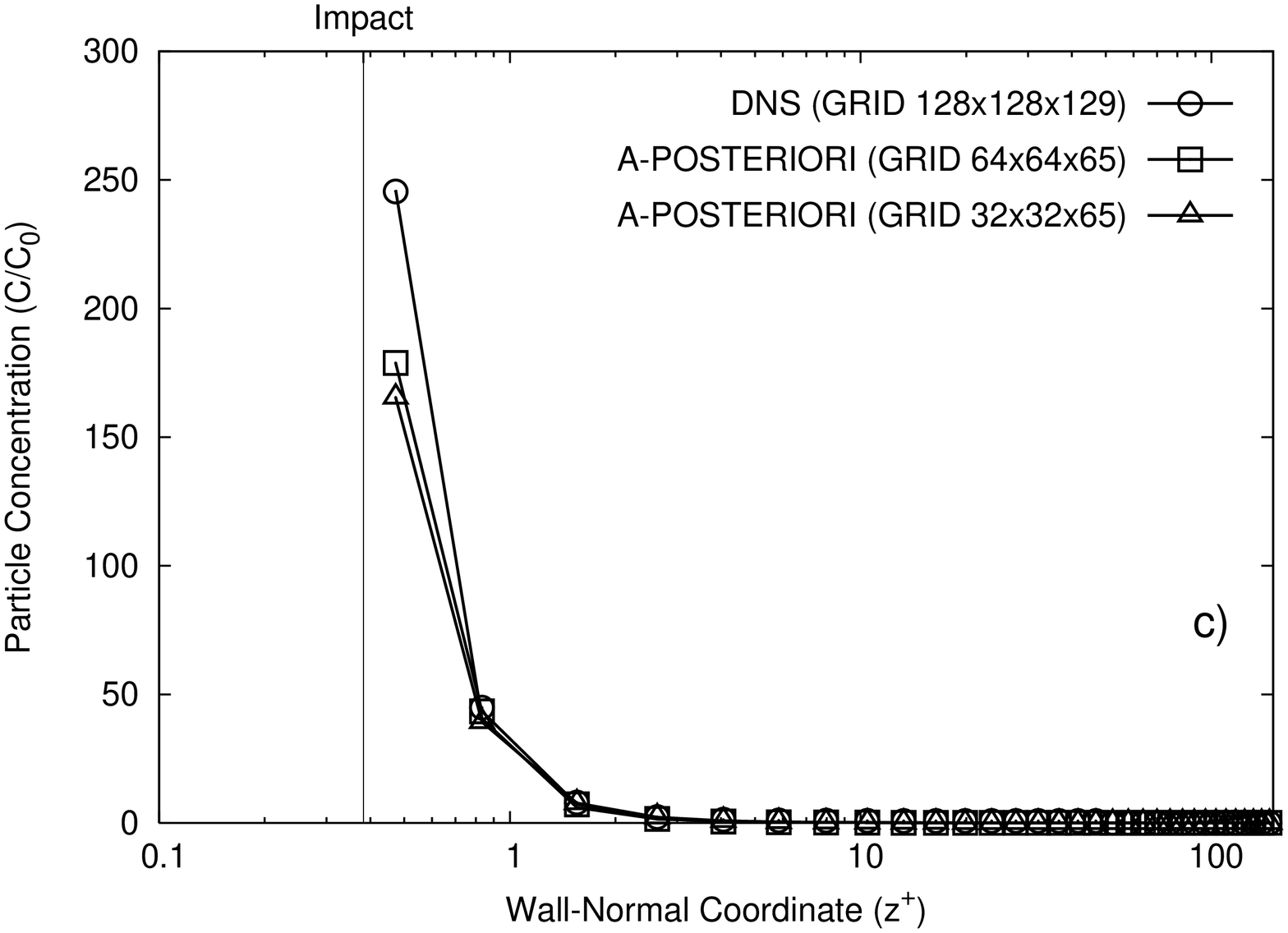}}
\vspace{0.2cm}
\caption{
Particle concentration in {\em a-posteriori} tests without SGS modeling
in the particle equation of motion: comparison between DNS ($\bigcirc$),
{\em a-posteriori} LES on the fine $64 \times 64 \times 65$ grid ($\square$)
and {\em a-posteriori} LES on the coarse $32 \times 32 \times 65$ grid ($\triangle$).
Panels: (a) $St=1$ particles, (b) $St=5$ particles, (c) $St=25$ particles.
The vertical solid line in each
diagram indicates the position where the particles hit the wall
({\em Impact}): note that impact for the $St=1$ particles occurs
at $z^+=0.034$, outside the $z^+$-range covered in panel (a).
}
\label{a-posteriori-filtering}
\end{figure}

%
%

\clearpage
\newpage

\begin{figure}
\centerline{\includegraphics[height=12.0cm,angle=270]{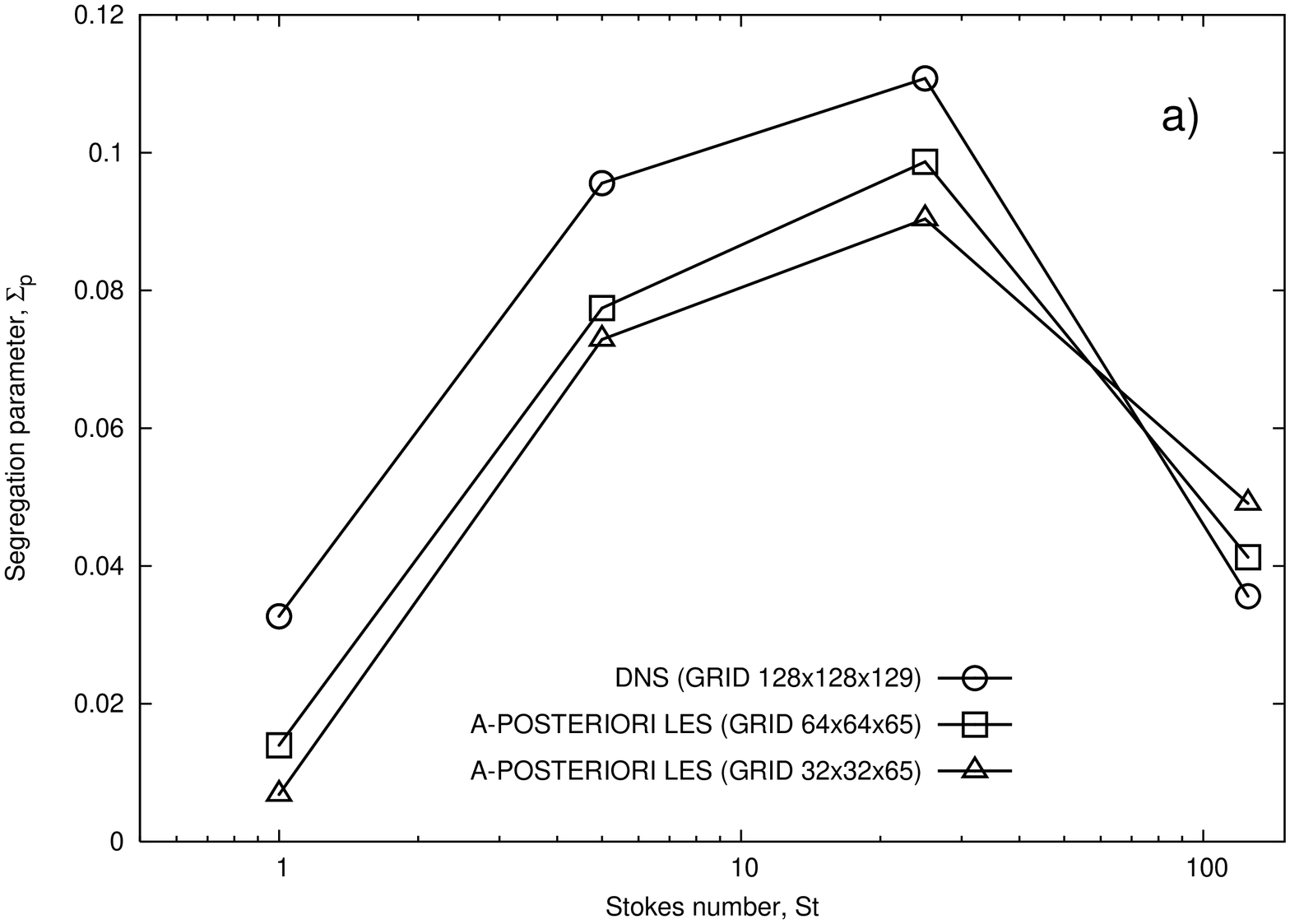}}
\centerline{\includegraphics[height=12.0cm,angle=270]{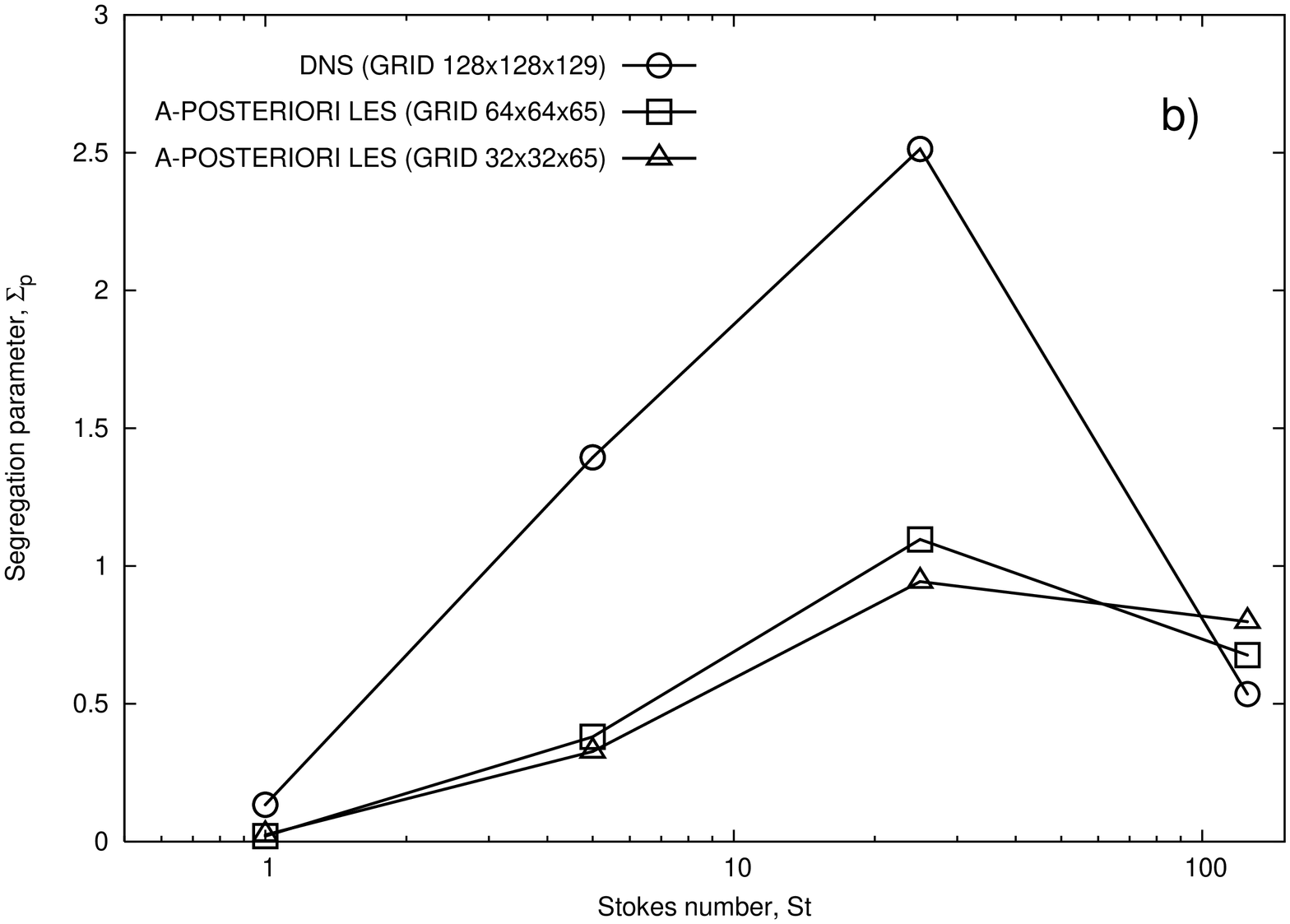}}
\vspace{0.5cm}
\caption{
Particle segregation, $\Sigma_p$, versus particle Stokes number, $St$,
in turbulent channel flow: comparison between DNS ($\bigcirc$),
{\em a-posteriori} LES on the fine $64 \times 64 \times 65$ grid ($\square$)
and {\em a-posteriori} LES on the coarse $32 \times 32 \times 65$ grid ($\triangle$).
Panels: (a) channel centerline ($145 \le z^+ \le 150$), (b) near-wall region ($0 \le z^+ \le 5$).
}
\label{poisson-dns-a-posteriori-les}
\end{figure}

%
%

\clearpage
\newpage

\begin{figure}
\centerline{\includegraphics[height=12.0cm,angle=270]{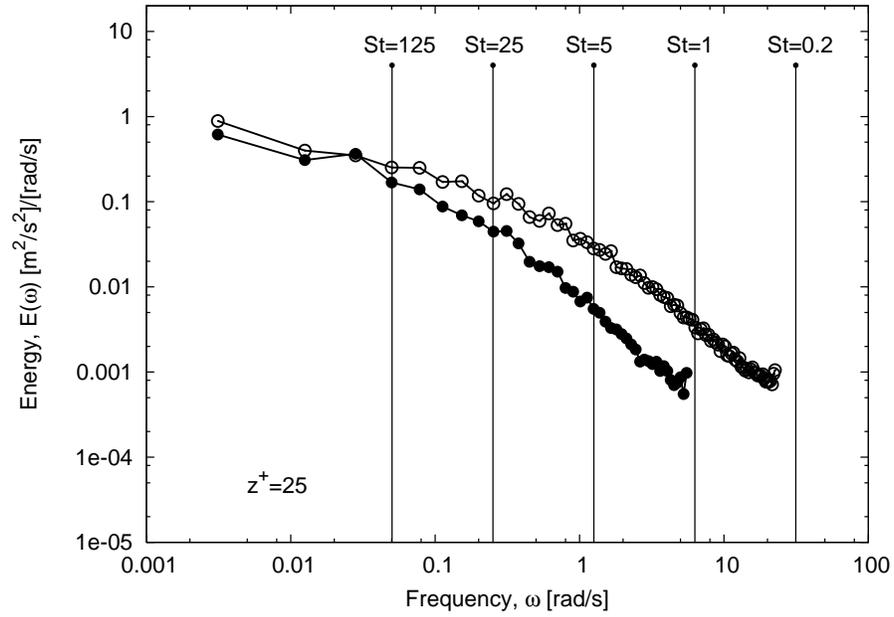}}
\vspace{0.2cm}
\caption{One-dimensional (streamwise) frequency spectrum for turbulent channel flow
computed at $z^+=25$ for two different Reynolds numbers: $Re_{\tau}^l=150$ ($\bullet$)
and $Re_{\tau}^h=300$ ($\circ$).}
\label{spectrum}
\end{figure}

%
%

\clearpage
\newpage

\begin{figure}[h]
\centering
\includegraphics[height=8.5cm,angle=270]{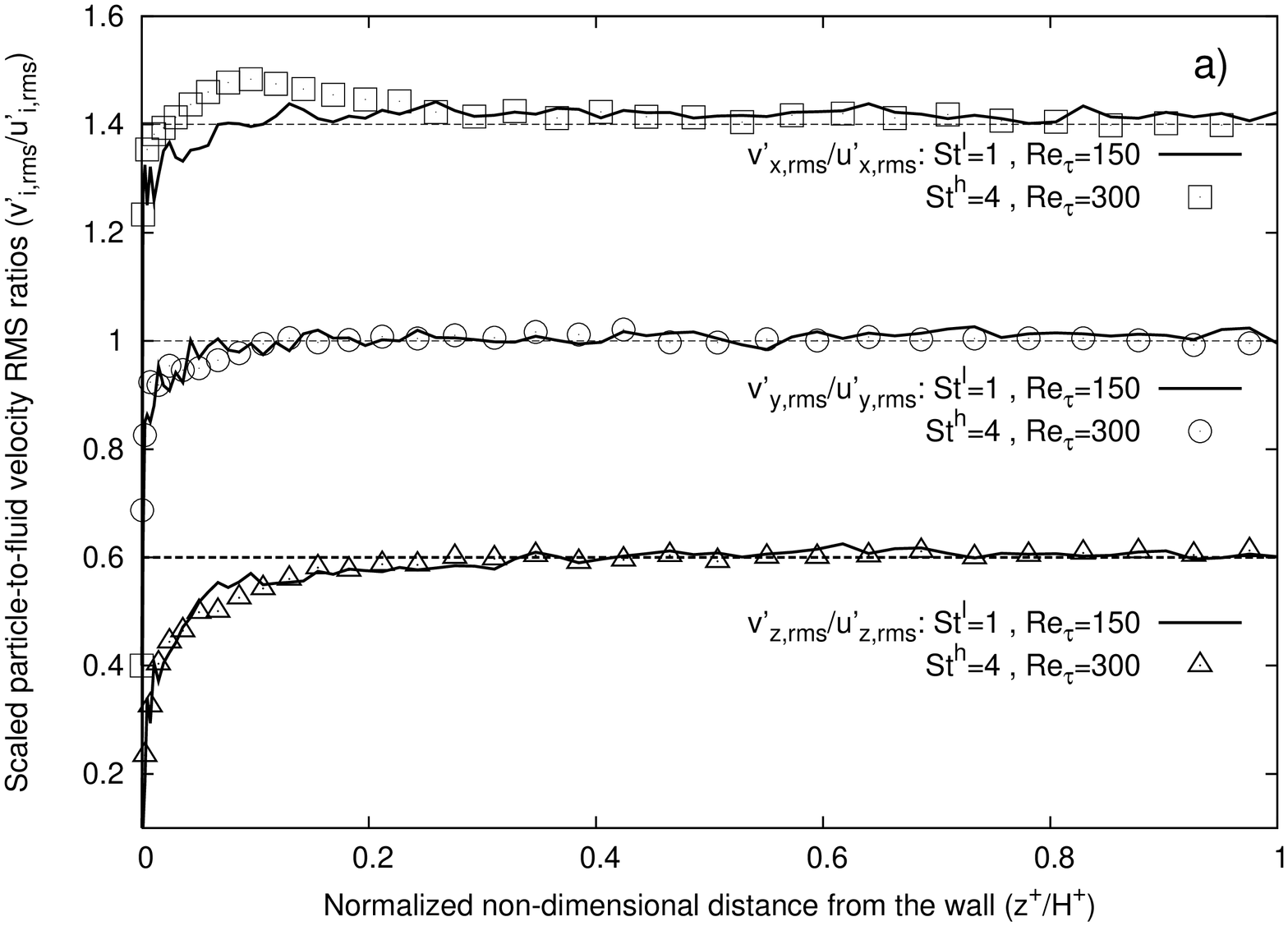}
\includegraphics[height=8.5cm,angle=270]{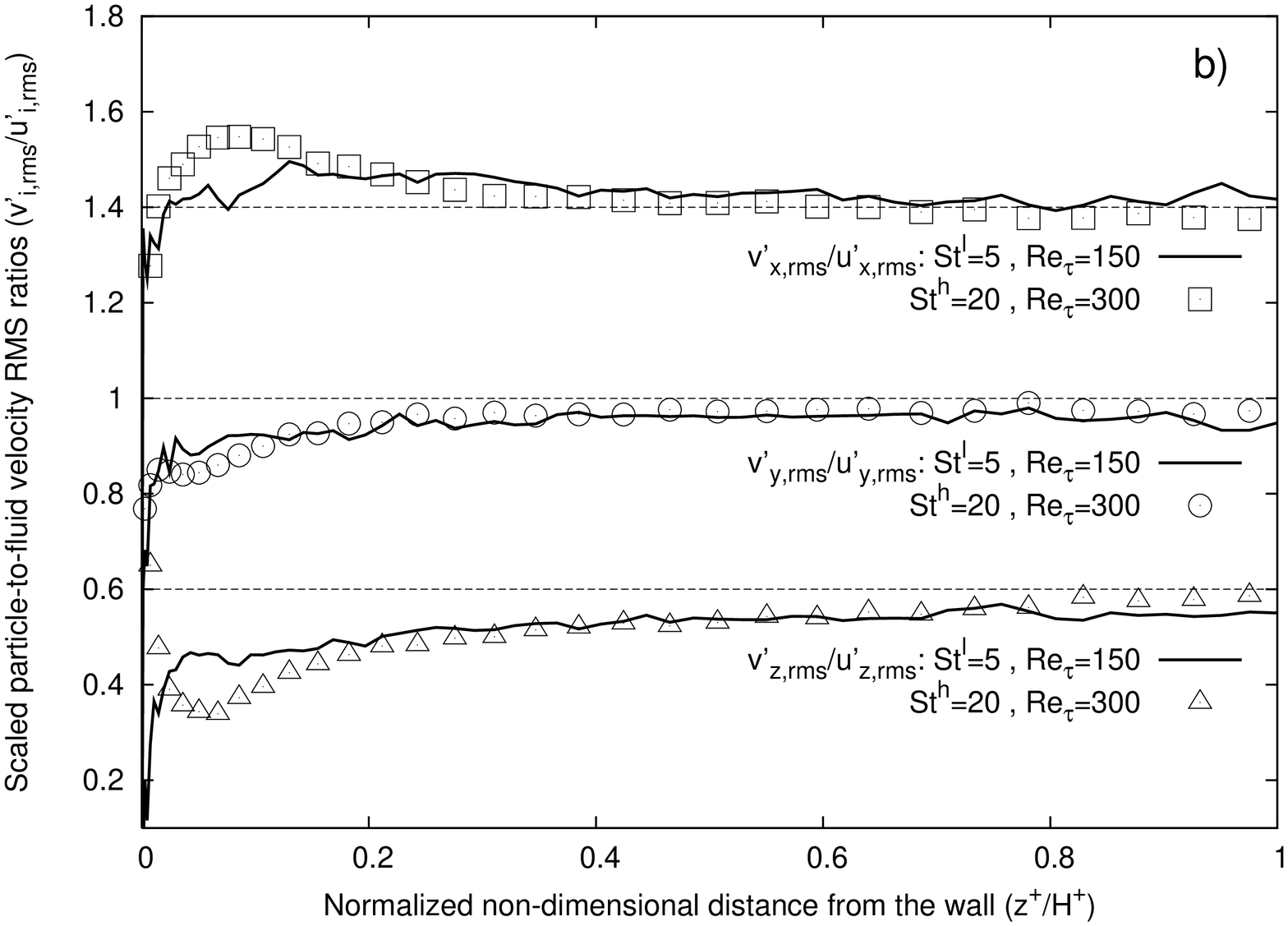}
\includegraphics[height=8.5cm,angle=270]{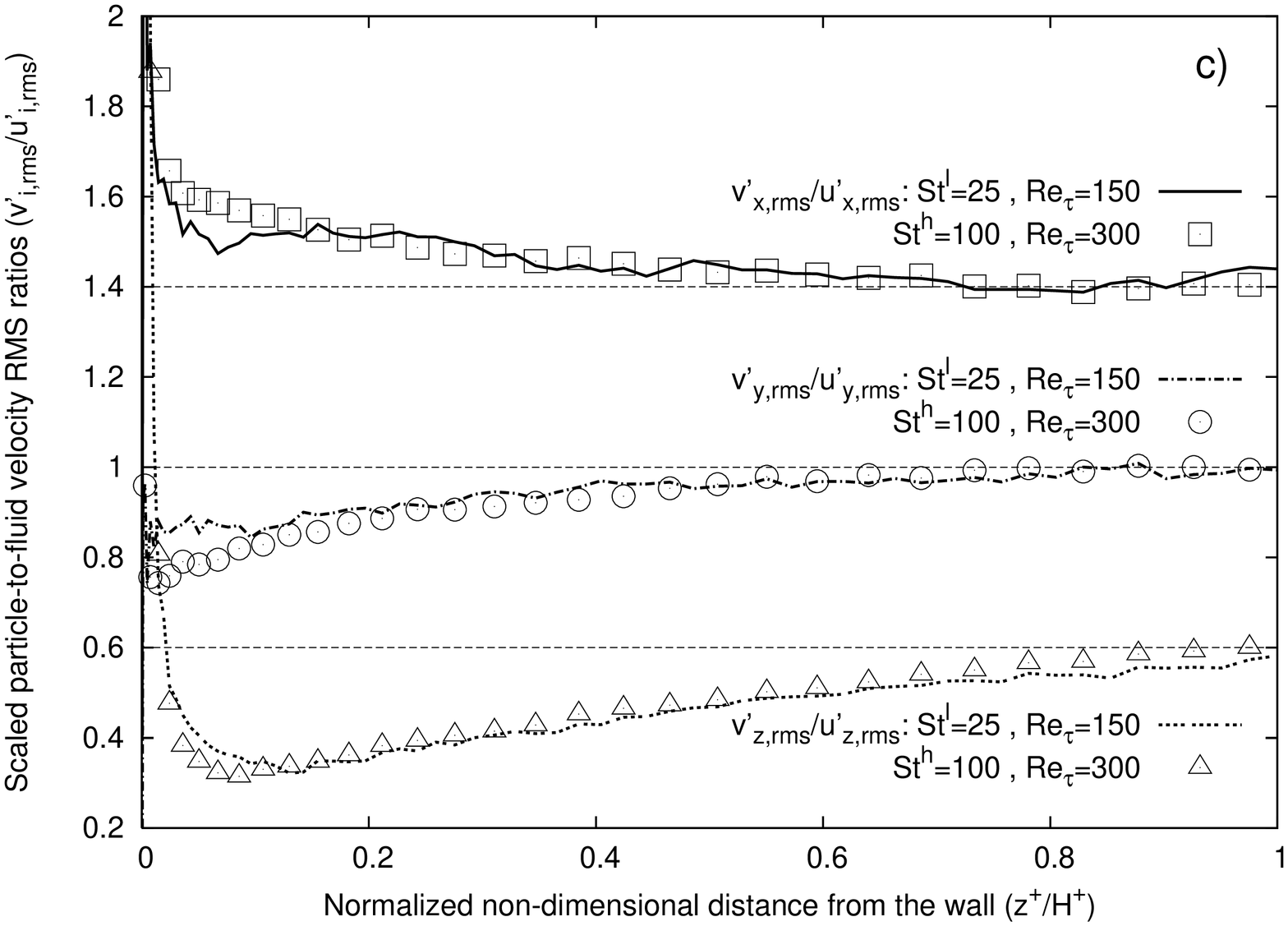}
\vspace{0.5cm}
\caption{
Scaled
particle$-$to$-$fluid
velocity rms ratios at low Reynolds number,
$( v'_{i,rms}/u'_{i,rms} ) |_{St^l, Re_{\tau}^l}$, and at high
Reynolds number, $( v'_{i,rms}/u'_{i,rms} ) |_{St^h, Re_{\tau}^h}$.
Panels: (a) $St^l=1$ versus $St^h=4$, (b) $St^l=5$ versus $St^h=20$, (c) $St^l=25$ versus $St^h=100$.
}
\label{scaling-DNS}
\end{figure}

%
%

\begin{figure}
\centerline{\includegraphics[height=8.0cm,angle=0]{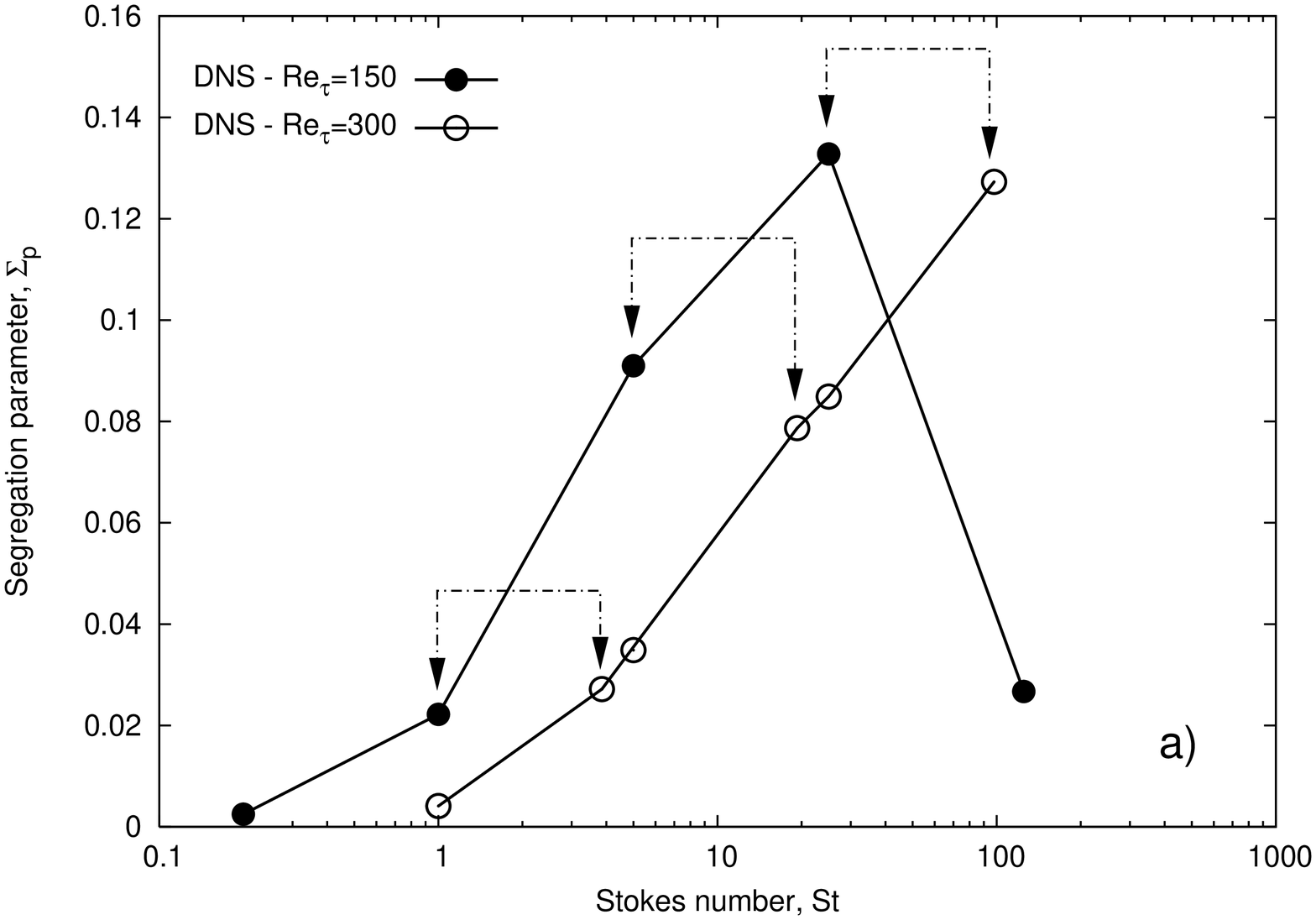}}
\centerline{\includegraphics[height=8.0cm,angle=0]{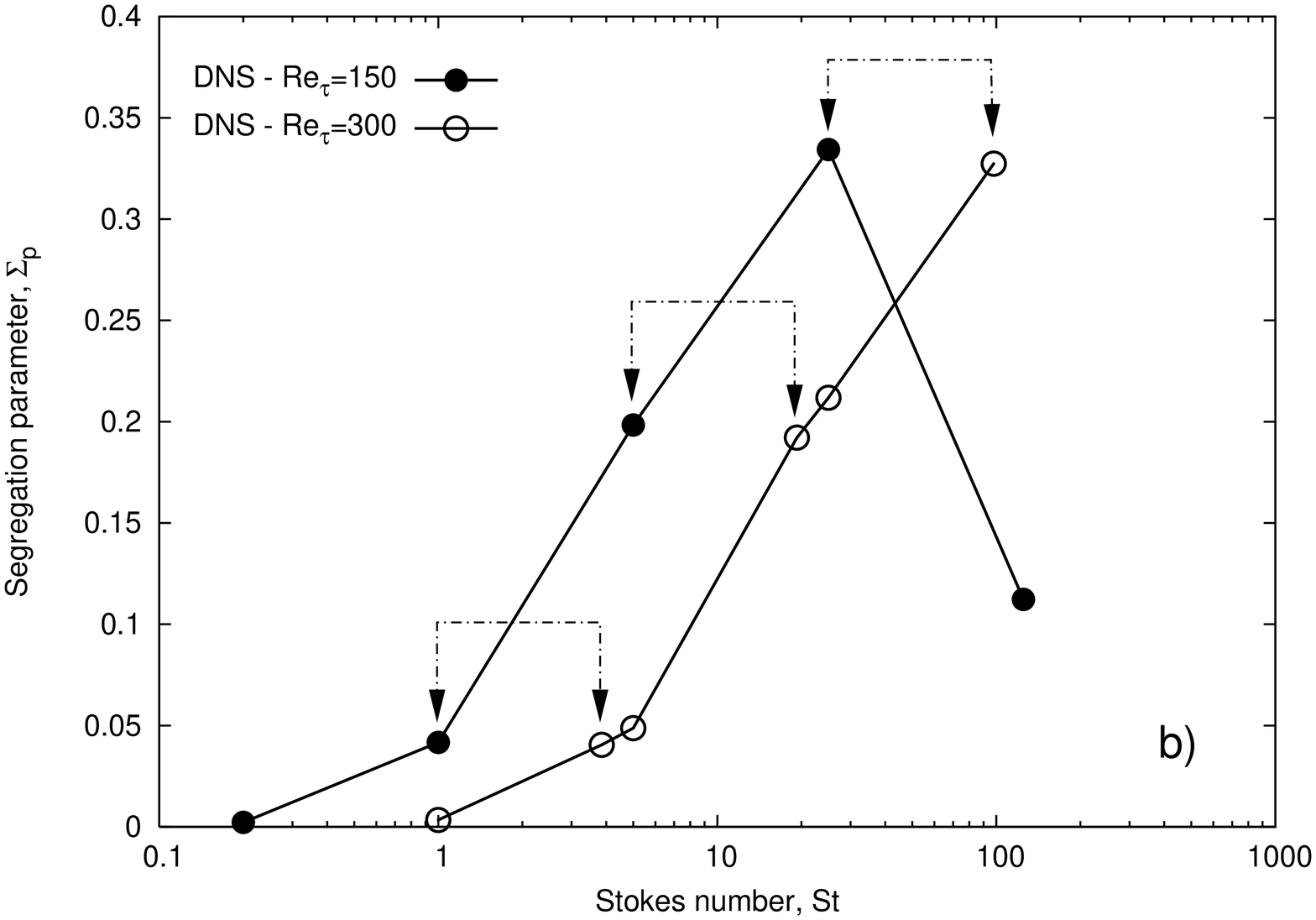}}
\vspace{0.5cm}
\caption{
Particle segregation parameter, $\Sigma_p$, versus particle Stokes number, $St$,
in turbulent channel flow at two different Reynolds numbers: $Re_{\tau}=150$ ($\bullet$)
and $Re_{\tau}=300$ ($\circ$).
Panels: (a) channel centerline ($145 \le z^+ \le 150$), (b) near-wall region ($0 \le z^+ \le 5$).
}
\label{poisson-dns-dns}
\end{figure}

%% file: main.bbl
\begin{thebibliography}{xx}
\bibitem{footnote}
In the present flow configuration,
the non-dimensional Kolmogorov length scale, $\eta_K^+$,
varies along the wall-normal direction from a minimum
value $\eta_K^+=1.6$ at the wall to a maximum value
$\eta_K^+=3.6$ at the centerline. The grid resolution
in the wall-normal direction is such that the
first collocation point is at $z^+ = 0.05$ from the wall,
while in the center of the channel $\Delta z^+=3.7$
\end{thebibliography}
